\documentclass[12pt]{iopart}
\newtheorem{thm}{Theorem}
\newtheorem{lem}{Lemma}
\newtheorem{pro}{Proposition}
\newtheorem{con}{Conjecture}

\usepackage[dvips]{graphicx}
\usepackage{iopams}
\begin{document}

\title[On integrals for some class of ordinary difference equations 
]{On integrals for some class of ordinary difference equations 
admitting a Lax pair representation
}

\author{Andrei K Svinin}

\address{Institute for System
Dynamics and Control Theory, Siberian Branch of
Russian Academy of Sciences, Russia}
\ead{svinin@icc.ru}
\begin{abstract}
We consider two infinite classes of ordinary difference equations admitting Lax pair representation. Discrete equations in these classes are parameterized  by two integers $k\geq 0$ and $s\geq k+1$. We describe the first integrals for these two classes  in terms of special discrete polynomials. We show an equivalence of two difference equations belonged to different classes corresponding to the same  pair $(k, s)$. We show that solution spaces $\mathcal{N}^k_s$ of different ordinary difference equations with fixed value of $s+k$ are organized in chain of inclusions. 
\end{abstract}

\vspace{2pc}


\submitto{J. Phys. A: Math. Theor.}

\section{Introduction}
\label{sect1}

In  this paper we consider  some infinite classes of $N$th order autonomous ordinary difference equations yielded by nonlinear recurrence relations
\begin{equation}
T(i+N)=R(T(i),\ldots, T(i+N-1)),\;\;i\in\mathbb{Z}
\label{1}
\end{equation}
with  right-hand sides being some rational functions of their arguments. Every equation of the form (\ref{1}) generate a special birational map $\mathbb{R}^N\mapsto\mathbb{R}^N$ for real-valued initial data $\left\{y_j\equiv T(i_0+j) : j=0,\ldots, N-1\right\}$ 
\[
\left(y_0,\ldots, y_{N-1}\right)\mapsto \left(y_1,\ldots, y_{N-1}, F(y_0,\ldots, y_{N-1})\right)
\]  
provided that one may rewrite (\ref{1}) in equivalent form as 
\[
T(i-1)=\bar{R}(T(i),\ldots, T(i+N-1))
\]
with corresponding rational function $\bar{R}$ of its arguments.


It should be noted that in a sense ordinary difference equations (\ref{1}) are natural discrete counterparts of their differential analogs and many of the notions encountered in the theory of ordinary differential equations  can be extended to them.  For instance, one says that the function $J=J(i)=J(T(i),\ldots, T(i+N-1))$ is a first integral for (\ref{1}) if by virtue of this equation one has $J(i+1)=J(i)$. It is very important that the notion of Liouville-Arnold integrability also can  be extended to symplectic maps  with suitable modification of Liouville theorem \cite{Maeda}, \cite{Bruschi}, \cite{Veselov}. Many examples of integrable autonomous ordinary difference equations were given, for example,  in \cite{Kamp}, \cite{Kamp1}, \cite{Tran2}, \cite{Tran3} etc. Experience suggests that such equations are usually grouped in infinite classes of their own kind and their first integrals have a similar appearance.

A goal that we set in this paper is to describe the first integrals for some classes of difference equations (\ref{1}) in terms of their associated discrete polynomials. Specifically, we consider ordinary difference equations that admit a Lax representation and might be integrable in a Liouville-Arnold sense. In addition, the peculiar property of these difference equations and their Lax pairs  is that they are presented in terms of special discrete polynomials \cite{Svinin6}.  This suggests that their first integrals also should be expressed via these polynomials.  The article aims to address the problem of constructing of the first integrals for whole classes of ordinary difference equations, which are derived from Lax representation.
In the following section we first describe discrete $(h, n)$-polynomials of one field $T=T(i)$ and  associate to them the ordinary difference equations. These classes of polynomials were introduced in \cite{Svinin1} and \cite{Svinin6}. They are closely related to formal pseudo-difference operators. Also in this section, we show  a couple of examples of ordinary difference equations corresponding to the discrete $(1, 2)$-polynomials  and provide the reader with the first integrals  derived from the Lax representation. It should be noted here that, in general,   there is no problem to construct first integrals for ordinary difference equation under consideration from corresponding the Lax representation. The problem we set in the article is to construct the first integrals for whole classes of discrete equations and reveal the relationship of these integrals with corresponding discrete polynomials. Unfortunately, only in the case $(h, n)=(1, 1)$ we have advanced enough in addressing this problem. In section \ref{sect3}, we discuss the discrete $(1, 1)$-polynomials to prepare the ground for further investigation. In section \ref{sect4}, we discuss the Lax representation for two infinite classes of ordinary difference equations in the case $(h, n)=(1, 1)$. The equations in these  classes are parametrized by pairs of integers $k\geq 0$ and $s\geq k+1$.  Based on actual calculations using Lax representation,  in section \ref{sect5}, we construct  the first integrals starting from in  a sense universal discrete polynomial (rational function) which yields all the first integrals which can be obtained from the Lax representation for all equations in the corresponding class. With some technical conjecture, we show that two $(k, s)$th discrete equations with any fixed $(k, s)$ belonging to different, at the first glance, classes, are related to each other through interchange of integrals and parameters (cf. \cite{Roberts}). 
Also we prove that solution spaces $\mathcal{N}^k_s$ with fixed values of the sum $s+k$ are organized in chains of inclusions.

\section{Discrete $(h, n)$-polynomials and ordinary difference equations}
\label{sect2}

To begin with, we fix some notion which will be used in the sequel. Throughout the paper we call any polynomial $P=P(T(i+a),\ldots, T(i+b))$ of its arguments with $a\leq b$ a discrete polynomial of the field $T=T(i)$. This notion is in analogy with the notion of differential polynomial. We use also  discrete rational functions which can be presented as a ratio of two discrete polynomials.

\subsection{Ordinary difference equations}

In \cite{Svinin6} we have shown a large number of  systems of ordinary difference equations  which have a Lax pair representation. In particular, two $(h, n)$-classes of one-field systems in this  collection are presented by    
\begin{equation}
T(i+sh+kn)\tilde{T}^k_s(i)=T(i)\tilde{T}^k_s(i+h+n)
\label{12}
\end{equation}
and
\begin{equation}
T(i+kn)\tilde{S}^k_s(i)=T(i+sh)\tilde{S}^k_s(i+h+n).
\label{13}
\end{equation}
$\tilde{T}^k_s(i)$ and $\tilde{S}^k_s(i)$  in (\ref{12}) and (\ref{13}), respectively,  are non-homogeneous discrete polynomials of the field $T=T(i)$ to be described. 

\subsection{Discrete polynomials $\tilde{T}^r_s(i)$}

To begin with, we first describe polynomials $\tilde{T}^r_s(i)$ in this subsection  and then we define polynomials $\tilde{S}^r_s(i)$. Let  $(h, n)$  be a pair of positive integers which are supposed to be co-prime. Let us define an infinite class of homogeneous discrete polynomials $\left\{T^r_s(i)\right\}$ via the explicit formula\footnote{Strictly speaking, we should indicate a dependence for each class of polynomials on $h$ and $n$, but we prefer to use simplified notations for these polynomials in the hope that it will not lead to a confusion.}
\begin{equation}
T^r_s(i):=\sum_{\{\lambda_j\}\in D_{r, s}} T(i+\lambda_1h)T(i+\lambda_2h+n)\cdots T(i+\lambda_rh+(r-1)n)
\label{exp}
\end{equation}
with $D_{r, s}:= \{\lambda_j : 0\leq \lambda_1<\cdots<\lambda_r\leq s-1\}$, for $r\geq 0$ and $s\geq 1$. In particular, we put $T^0_s(i)\equiv 1$.
It is evident that, from the definition, it follows that 
\begin{equation}
T^r_s(i)\equiv 0,\;\forall\; s\leq r-1
\label{Tks}
\end{equation}
since in this case $D_{r, s}=\emptyset$. 
Polynomials (\ref{exp}), by virtue of their definition, satisfy identities \cite{Svinin6}
\begin{eqnarray}
T^r_s(i)&=&T^r_{s-1}(i+h)+T(i)T^{r-1}_{s-1}(i+h+n) \label{t1} \\
&=&T^r_{s-1}(i)+T(i+(s-1)h+(r-1)n)T^{r-1}_{s-1}(i).  \label{t2}
\end{eqnarray}
Indeed, it can be seen that identity (\ref{t1})  corresponds to the partition of $D_{r, s}$  into two non-intersecting sets  
\[
D^{(1)}_{r, s}:=\{\lambda_j : \lambda_1=0,\;\;  1\leq\lambda_2<\cdots<\lambda_r\leq s-1\}
\]
and
\[
D^{(2)}_{r, s}:=\{\lambda_j : 1\leq\lambda_1<\cdots<\lambda_r\leq s-1\},
\]
while identity (\ref{t2}) appears as a result of the partition  $D_{r, s}=\tilde{D}^{(1)}_{r, s}\sqcup  \tilde{D}^{(2)}_{r, s}$ with two parts
\[
\tilde{D}^{(1)}_{r, s}:=\{\lambda_j : 0\leq\lambda_1<\cdots<\lambda_{r-1}\leq s-2,\;\;  \lambda_r=s-1\}
\]
and
\[
\tilde{D}^{(2)}_{r, s}:=\{\lambda_j : 0\leq\lambda_1<\cdots<\lambda_r\leq s-2\}.
\]
Taking as initial conditions the data $T^0_s(i)\equiv 1,\; \forall s\geq 1$ and $T^1_1(i)=T(i),\; T^r_1(i)\equiv 0,\; \forall r\geq 2$  
and using  then one of recurrent relations (\ref{t1}) or (\ref{t2}), we ultimately obtain the whole $(h, n)$-class of homogeneous discrete polynomials $\{T^r_s(i)\}$. 


It is a simple observation that discrete polynomials $T^{r-jh}_{s+jn}(i)$, for any fixed $j=0,\ldots, \kappa$, satisfies recurrent relations of the same form as  (\ref{t1}) and (\ref{t2}), 
where  $r=\kappa h+\bar{r}$, i.e.,  $\bar{r}$ is supposed to be the remainder of division of $r$ by $h$.
This means that, in general, we can present the common solution of  (\ref{t1}) and (\ref{t2}) as a linear combination\footnote{It is evident that in equation (\ref{12}), without loss of generality, we can put $c_0=1$.}
\begin{equation}
\tilde{T}^r_s(i):= \sum_{j=0}^{\kappa}c_jT^{r-jh}_{s+jn}(i),
\label{non}
\end{equation}
where $c_j$ are arbitrary constants.
Remark that, due to (\ref{Tks}), it makes sense to consider non-homogeneous discrete polynomials (\ref{non}) and consequently discrete equations of the form (\ref{12}) only for  $s\geq k$. Moreover, since
\[
T^k_k(i)=\prod_{j=1}^kT(i+(j-1)(h+n))
\]
and  therefore the relation
\[
T(i+k(h+n))T^k_k(i)=T(i)T^k_k(i+h+n)
\]
is an identity, then equation (\ref{12}) in the case $s=k$ is reduced to the following one:
\[
T(i+k(h+n))\tilde{T}^{k-h}_{k+n}(i)=T(i)\tilde{T}^{k-h}_{k+n}(i+h+n).
\]
This means that, without loss of generality, we may consider an $(h, n)$-class of the ordinary difference equations of the form (\ref{12}) only for $s\geq k+1$.

\subsection{Discrete polynomials $\tilde{S}^r_s(i)$}

Non-homogeneous discrete polynomials  $\tilde{S}^r_s(i)$ are given by  linear combinations\footnote{In what follows, we assume that $H_0=1$.}
\begin{equation}
\tilde{S}^r_s(i):= \sum_{j=0}^{\kappa}(-1)^{jh}H_jS^{r-jh}_{s-jn}(i+jnh),
\label{non1}
\end{equation}
where $H_j$  are supposed to be arbitrary constants. It is supposed in (\ref{non1}) that $s-jn\leq 1$ and $r-jh\leq 0$. The homogeneous discrete polynomials $S^{k}_{s}(i)$ can be defined explicitly as
\[
S^r_s(i):=\sum_{\{\lambda_j\}\in B_{r, s}} T(i+\lambda_1h+(k-1)n)\cdots T(i+\lambda_{r-1}h+n)T(i+\lambda_rh),
\]
where the summation is performed over the set $B_{r, s}\equiv \{\lambda_j : 0\leq \lambda_1\leq\cdots\leq\lambda_r\leq s-1\}$. These polynomials satisfy identities 
\begin{eqnarray}
S^r_s(i)&=&S^r_{s-1}(i+h)+T(i+(r-1)n)S^{r-1}_{s}(i) \label{s11} \\
&=&S^r_{s-1}(i)+T(i+(s-1)h)S^{r-1}_{s}(i+n)  \label{s21}
\end{eqnarray}
which come from the partitions $B_{r, s}=B^{(1)}_{r, s}\sqcup  B^{(2)}_{r, s}$ and $B_{r, s}=\tilde{B}^{(1)}_{r, s}\sqcup  \tilde{B}^{(2)}_{r, s}$, respectively, with 
\[
B^{(1)}_{r, s}:=\{\lambda_j :\lambda_1=0,\;\;  0\leq\lambda_2\leq\cdots\leq\lambda_r\leq s-1\},
\]
\[
B^{(2)}_{r, s}:=\{\lambda_j : 1\leq\lambda_1\leq\cdots\leq\lambda_r\leq s-1\}
\]
and
\[
\tilde{B}^{(1)}_{r, s}:=\{\lambda_j : 0\leq\lambda_1\leq\cdots\leq\lambda_{r-1}\leq s-1,\;\;  \lambda_r=s-1\},
\]
\[
\tilde{B}^{(2)}_{r, s}:=\{\lambda_j : 0\leq\lambda_1\leq\cdots\leq\lambda_r\leq s-2\}.
\]
Taking as initial conditions the data $S^0_s(i)\equiv 1,\; \forall s\geq 1$ and 
\[
S^r_1(i)=T(i)T(i+n)\cdots T(i+(r-1)n),\;  \forall r\geq 1
\]  
and using  one of recurrent relations (\ref{s11}) or (\ref{s21}), we  obtain the $(h, n)$-class of homogeneous discrete polynomials $\{S^r_s(i)\}$.  It can be checked that the polynomials $S^{r-jh}_{s-jn}(i+jnh)$ and their linear combinations of the form (\ref{non1}) satisfy (\ref{s11}) and (\ref{s21}).
\begin{lem} \label{lem0} 
The sets of homogeneous discrete polynomials $T:=\{T^r_s(i)\}$ and $S:=\{S^r_s(i)\}$ are related to each other through identities \cite{Svinin6}
\begin{equation}
\left(S, T\right)^r_s(i):= \sum_{j=0}^{r}(-1)^jS^{r-j}_s(i)T^{j}_s(i+(r-j)n)=0,\;\; \forall r, s\geq 1
\label{ST2}
\end{equation}
and
\begin{equation}
\left(T, S\right)^r_s(i):= \sum_{j=0}^{r}(-1)^jT^{r-j}_s(i)S^{j}_s(i+(r-j)n)=0,\;\; \forall r, s\geq 1.
\label{ST1}
\end{equation}
\end{lem}
We place the proof of this lemma based on set-theoretical reasonings in Appendix. It should be noted that this proof in principle do not depend on $(h, n)$. 


For (\ref{ST2}) and (\ref{ST1}), let us make some remarks. Let us assign to each set of discrete polynomials $F=\{F^r_s(i) : s\geq 1, r\geq 1\}$ such that $F^0_s(i)\equiv 1$, semi-infinite  lower triangular matrices $\mathrm{F}_s(i)$ with matrix elements 
\[
(\mathrm{F}_s(i))_{pq}=F^{p-q}_s(i+(q-1)n),\; p\leq q.
\]  
	As can be easily checked, a product $(\mathrm{F}, \mathrm{G})_s(i):=D\mathrm{F}_s(i)D\mathrm{G}_s(i)$, where $D:=\mathrm{diag}(1, -1, 1,\ldots)$ has matrix elements $\left(G, F\right)^{p-q}_s(i+(q-1)n)$. This means that we can represent an infinite number of identities    (\ref{ST1}) in the form of matrix relations:  
\begin{equation}
(\mathrm{T}, \mathrm{S})_s(i)\equiv E,\; \forall s\geq 1,
\label{TS}
\end{equation}
where $E$ stands for the unit semi-infinite matrix. Since  any semi-infinite  lower triangular matrix has its unique inverse, one can easily derive that from (\ref{TS}) it follows  that $(\mathrm{S}, \mathrm{T})_s(i)\equiv E$ and vice versa.
Equivalently, this means that (\ref{ST2}) implies  (\ref{ST1}) and vice versa. Also, we remark that matrix representation of identities (\ref{ST2}) and (\ref{ST1}) makes it possible to express all the discrete polynomials  $S^r_s(i)$ via $T^r_s(i)$ in determinant form and vice versa.

\subsection{Equivalent forms of discrete equations (\ref{12}) and (\ref{13})}

We remark that both equations (\ref{12}) and (\ref{13}) are autonomous ordinary difference equations of the form of nonlinear recurrence relation (\ref{1}) with $N=sh+kn$. Substituting
\[
\tilde{T}^k_s(i)=\tilde{T}^k_{s-1}(i+h)+T(i)\tilde{T}^{k-1}_{s-1}(i+h+n) 
\]
into the left-hand side of (\ref{12}) and 
\[
\tilde{T}^k_s(i+n+h)=\tilde{T}^k_{s-1}(i+n+h)+T(i+kn+sh)\tilde{T}^{k-1}_{s-1}(i+n+h) 
\]
into the right-hand side of this equation we obtain it in an equivalent form
\begin{equation}
T(i+kn+sh)\tilde{T}^k_{s-1}(i+h)=T(i)\tilde{T}^k_{s-1}(i+n+h).
\label{14}
\end{equation}
By analogy, we find out that equation (\ref{13}) is equivalent to
\begin{equation}
T(i+kn)\tilde{S}^k_{s+1}(i)=T(i+sh)\tilde{S}^k_{s+1}(i+n).
\label{15}
\end{equation}

\subsection{Lax pair representations}

Equations of the form (\ref{12}) and (\ref{13}) share the property of having Lax pair representation. The first linear equation is common for them, namely,   
\begin{equation}
z\psi_{i+n}+T(i+n)\psi_{i}=z\psi_{i+h+n}.
\label{155}
\end{equation}
$(k, s)$th equation in $(h, n)$-class of equations (\ref{12}) arise as a compatibility condition of (\ref{155}) and linear equation \cite{Svinin6}
\begin{equation}
w\psi_{i+sh+kn}=\sum_{j=0}^kz^{k-j}\tilde{T}^j_{s}(i+(k-j+1)n)\psi_{i+(k-j)n}.
\label{156}
\end{equation}
In turn, $(k, s)$th equation in an $(h, n)$-class of equations (\ref{13}) appears as a compatibility condition of (\ref{155}) and
\begin{equation}
w\psi_{i-sh+kn}=\sum_{j=0}^kz^{k-j}(-1)^j\tilde{S}^j_s(i-(j-k-1)n-sh)\psi_{i+(k-j)n}.
\label{157}
\end{equation}

The following remarks are in order. It is quite natural to suppose that ordinary difference equations admitting a Lax pair representation should be integrable in Liouville-Arnold sense and proving this fact is the general problem. We can derive some number of first integrals for ordinary difference equation under consideration making use of its Lax pair, but, unfortunately,  this do not provide us the full information about the integrability of the equation under consideration since we do not know their Poisson structure. Moreover, in general, we do not know whether a full number of integrals is obtained from corresponding Lax pair. 

 To extract the first integrals of the equation under consideration, we rewrite the corresponding two linear equations
in equivalent matrix form 
\begin{equation}
L_i\Psi_i=0,\;\; \Psi_{i+1}=A_i\Psi_i
\label{Lax pair}
\end{equation}
where $\Psi_i:=\left(\psi^{(0)}_i,\ldots, \psi^{(\mathcal{N}-1)}_i\right)^T$, with $\psi^{(j)}_i:=\psi_{i+j}$ and $\mathcal{N}=sh+kn$ for linear problem (\ref{156}) and 
\[
\mathcal{N}=\left\{
\begin{array}{cc}
sh,\;\;\mbox{if}\;\; sh-kn\geq 0, \\
kn,\;\;\mbox{if}\;\; sh-kn< 0,
\end{array}
\right.
\]
for (\ref{157}). Clearly, $L=L_i$  in (\ref{Lax pair}) is some $\mathcal{N}\times \mathcal{N}$ matrix-function. The condition $\det(L)=0$ gives some polynomial equation $P(w,z)=0$, yielding corresponding spectral curve and first integrals. Below we give some simple examples of ordinary difference equations of the form (\ref{12})  and (\ref{13}) together with their first integrals.  

\subsection{Examples}

Here we show two examples of ordinary difference equations (\ref{12}) and (\ref{13}) in the case $(h, n)=(1, 2)$. The case $(h, n)=(1, 1)$ is investigated in much greater detail in sections \ref{sect4} and \ref{sect5}.

\subsubsection{Equation of the form (\ref{12}) in the  case $(h, n)=(1, 2),\; k=1,\; s=2$}

Lax pair in this case is given by
\begin{equation}
z\psi_{i+3}=z\psi_{i+2}+T(i+2)\psi_i,\;\;
w\psi_{i+4}=z\psi_{i+2}+\tilde{T}^1_2(i+2)\psi_i.
\label{h=1, n=2, k=1, s=2}
\end{equation}
Ordinary difference equation (\ref{12}) in this case  becomes the fourth-order one: 
\[
T(i+4)=T(i)\frac{T(i+3)+c_1}{T(i+1)+c_1}.
\]
Using (\ref{h=1, n=2, k=1, s=2}), we obtain the following first integrals for this equation:\footnote{It is more convenient to present the first integrals in terms of  initial data $\{y_j\}$.} 
\[
\mathcal{H}_0=\frac{y_0y_1y_2y_3}{(y_1+c_1)(y_2+c_1)},
\]
\[
H_1=y_0+y_1+y_2+y_3+\frac{y_0y_3\left(y_1+y_2+c_1\right)}{(y_1+c_1)(y_2+c_1)},
\]
\[
G=\frac{y_0y_2}{y_1+c_1}+\frac{y_1y_3}{y_2+c_1}.
\]
Calculations show that a spectral curve in this case  is given by
\[
\mathcal{H}_0w^3+z^3\left(z+H_1\right)w-z^4\left(z+\tilde{H}_1\right)-Gz^2\left(w-z\right)^2=0
\]
with $\tilde{H}_1\equiv H_1+c_1$. Maple says that this algebraic curve is elliptic.

\subsubsection{Equation of the form (\ref{13}) in the  case $(h, n)=(1, 2),\; k=1,\; s=3$}

Lax pair in this case is given by
\begin{equation}
z\psi_{i+3}=z\psi_{i+2}+T(i+2)\psi_i,\;\;
w\psi_{i-1}=z\psi_{i+2}-\tilde{S}^1_3(i-1)\psi_i.
\end{equation}
Ordinary difference equation (\ref{13}) in this case looks as fifth-order one: 
\[
T(i+2)\left(T(i)+T(i+1)+T(i+2)-H_1\right)
\]
\[
\;\;=T(i+3)\left(T(i+3)+T(i+4)+T(i+5)-H_1\right).
\]
The Lax matrix $L$, in terms of initial data, looks, in this case, as
\[
\fl
L=\left(
\begin{array}{ccc}
\displaystyle
-\frac{y_2(y_2+y_3+y_4-H_1)}{w} & z&\displaystyle -z+\frac{zy_1}{w} \\
y_2-w&-y_0-y_1-y_2+H_1& z \\
w&y_0+y_1+y_2-H_1-w&-y_1-y_2-y_3+H_1
\end{array}
\right).
\]
From the latter, we obtain the following first integrals for this equation
\[
\mathcal{H}_0=y_2\left(y_0+y_1+y_2+y_3-H_1\right)\left(y_1+y_2+y_3+y_4-H_1\right),
\]
\[
H_2=-y_2\left(y_0+y_1+y_2\right)-y_3\left(y_1+y_2\right)-y_4y_2+y_2H_1,
\]
\[
\mathcal{H}_0\tilde{H}_1=-y_2\left(y_0+y_1+y_2-H_1\right)\left(y_1+y_2+y_3-H_1\right)\left(y_2+y_3+y_4-H_1\right).
\]
A spectral curve in this case  is given by
\[
zw^3-z\left(z+H_1\right)w^2-H_2zw-\mathcal{H}_0\left(z+\tilde{H}_1\right)=0.
\]
This curve is hyper-elliptic with genus equal 2. Performing the birational transformation
\[
w=Z,\;\;
z=\frac{1}{2}\frac{Z^3-H_1Z^2-H_2Z-\mathcal{H}_0-W}{Z^2},
\]
we get
\[
W^2=\left(Z^3-H_1Z^2-H_2Z-\mathcal{H}_0\right)^2-4\mathcal{H}_0\tilde{H}_1Z^2.
\]
 
\section{Discrete polynomials  in the case $(h, n)=(1, 1)$}
\label{sect3}

In this section we list a number of identities for discrete polynomials  $T^r_s(i)$ and  $S^r_s(i)$ in the case $(h, n)=(1, 1)$ that may be needed in the following. It should be noted that polynomials
$T^r_s(i)$ in this particular case first appeared in \cite{Kamp}. 

In addition, we introduce two associated classes of discrete polynomials $\{P^r_s(i)\}$ and $\{Q^r_s(i)\}$ which are necessary for the construction of the first integrals for corresponding ordinary difference  equations.


Clearly, identities (\ref{t1}) and  (\ref{t2}) are specified in this case as
\begin{eqnarray}
T^r_s(i)&=&T^r_{s-1}(i+1)+T(i)T^{r-1}_{s-1}(i+2) \label{t111} \\ 
&=&T^r_{s-1}(i)+T(i+s+r-2)T^{r-1}_{s-1}(i), \label{t222}
\end{eqnarray}
while (\ref{s11}) and  (\ref{s21}) become
\begin{eqnarray}
S^r_s(i)&=&S^r_{s-1}(i+1)+T(i+r-1)S^{r-1}_{s}(i) \label{s1} \\
&=&S^r_{s-1}(i)+T(i+s-1)S^{r-1}_{s}(i+1). \label{s2}
\end{eqnarray}
In what follows we will use identity (\ref{ST2}) which in the case $n=1$ case is specified as
\begin{equation}
\left(S, T\right)^r_s(i)= \sum_{j=0}^{r}(-1)^jS^{r-j}_s(i)T^{j}_s(i+r-j)=0.
\label{ST21}
\end{equation}

\subsection{Two associated classes of discrete polynomials $\{P^r_s(i)\}$ and $\{Q^r_s(i)\}$}

Here we define two classes of discrete polynomials which as it turns out are  necessary ingredients for the construction of the first integrals of ordinary difference equations under consideration. 

We first define a class of homogeneous discrete polynomials $\{P^r_s(i)\}$. To this aim, we consider the partition $D_{r, s}=\bar{D}_{r, s}\sqcup\tilde{D}_{r, s}$ with  
\[
\bar{D}_{r, s}:=\left\{\lambda_j : \lambda_1=0,\; 1\leq\lambda_2<\cdots<\lambda_{r-1}\leq s-2,\; \lambda_r=s-1\right\}
\]
and
\begin{eqnarray}
\tilde{D}_{r, s}&:=&\left\{\lambda_j : 1\leq\lambda_1<\cdots<\lambda_{r-1}\leq s-2,\; \lambda_r=s-1\right\}\nonumber\\
&&\sqcup\left\{\lambda_j : 0\leq\lambda_1<\cdots<\lambda_{r}\leq s-2\right\}\label{p1}\\
&=&\left\{\lambda_j : \lambda_1=0,\;\ 1\leq\lambda_2<\cdots<\lambda_{r}\leq s-2\right\}\nonumber\\
&&\sqcup\left\{\lambda_j : 1\leq\lambda_1<\cdots<\lambda_{r}\leq s-1\right\}.\label{p2}
\end{eqnarray}
Let us define a class of  discrete polynomials $\{P^r_s(i)\}$ by
\[
P^r_s(i)=\sum_{\{\lambda_j\}\in \tilde{D}_{r, s}}T(i+\lambda_1)T(i+\lambda_2+1)\cdots T(i+\lambda_r+r-1).
\]
It is important to notice the following. Since  $\tilde{D}_{r, s}=\emptyset$, for $s\leq r$, hence
\begin{equation}
P^r_s(i)\equiv 0,\;\;\mbox{for}\;\; s\leq r. 
\label{condition}
\end{equation}
We can easily find out the relationship between polynomials $P^r_s(i)$ and $T^r_s(i)$.  Since
\begin{eqnarray}
\fl
&&\sum_{\{\lambda_j\}\in \bar{D}_{r, s}}T(i+\lambda_1)T(i+\lambda_2+1)\cdots T(i+\lambda_r+r-1)\nonumber\\
\fl
&&\;\;\;\;\;\;\;\;=T(i)T(i+s+r-2)\sum_{1\leq\lambda_2<\cdots<\lambda_{r-1}\leq s-2}T(i+\lambda_2+1)\cdots T(i+\lambda_{r-1}+r-2)\nonumber\\ 
\fl
&&\;\;\;\;\;\;\;\;=T(i)T(i+s+r-2)T^{r-2}_{s-2}(i+2)\nonumber
\end{eqnarray}
then
\begin{equation}
P^r_s(i)= T^r_s(i)-T(i)T(i+s+r-2)T^{r-2}_{s-2}(i+2).
\label{s5}
\end{equation}
In turn, partitions (\ref{p1}) and (\ref{p2}) give the identities
\begin{eqnarray}
P^r_s(i)&=&T(i+s+r-2)T^{r-1}_{s-2}(i+1)+T^{r}_{s-1}(i)\label{PT1}\\
&=&T(i)T^{r-1}_{s-2}(i+2)+T^{r}_{s-1}(i+1),\label{PT2}
\end{eqnarray}
respectively.
Suppose now that  $T(i+s+r-1)=T(i)$. Taking into account (\ref{PT1}) and (\ref{PT2}), we obtain that by virtue of this periodicity condition
\begin{eqnarray}
P^r_s(i+1)&=&T(i+s+r-1)T^{r-1}_{s-2}(i+2)+T^{r}_{s-1}(i+1)\label{rep1}\\
&=&T(i)T^{r-1}_{s-2}(i+2)+T^{r}_{s-1}(i+1)\label{rep2}\\
&=&P^r_s(i).\nonumber
\end{eqnarray}
The latter shows the meaning of the discrete polynomials $P^r_s(i)$. They are the well-known first integrals of the periodicity equation $T(i+N)=T(i)$ (see, for example, \cite{Veselov1}) with $N=s+r-1$, while (\ref{PT1}) and (\ref{PT2}) give two expressions  of these integrals via polynomials $T^r_s(i)$. 


Now we define a class of discrete polynomials $\{Q^r_s(i)\}$, but previously let us prove  the following lemma. 
\begin{lem}
\label{lem1}
Two set-theoretic equalities
\begin{eqnarray}
B_{r, s+1}&\sqcup&\left\{\lambda_j : \lambda_1=-1,\; -1\leq\lambda_2\leq\cdots\leq\lambda_{r-1}\leq s,\; \lambda_r=s\right\}\nonumber \\
&=&B_{r, s}\sqcup\left\{\lambda_j :  -1\leq\lambda_1\leq\cdots\leq\lambda_{r-1}\leq s,\; \lambda_r=s\right\}\label{Bk}
\end{eqnarray}
and
\begin{eqnarray}
B_{r, s+1}^{-}&\sqcup&\left\{\lambda_j : \lambda_1=-1,\; -1\leq\lambda_2\leq\cdots\leq\lambda_{r-1}\leq s,\; \lambda_r=s\right\}\nonumber \\
&=&B_{r, s}\sqcup\left\{\lambda_j :\lambda_1=-1,\;  -1\leq\lambda_2\leq\cdots\leq\lambda_{r}\leq s\right\}\label{Bk1} 
\end{eqnarray}
are valid.
\end{lem}
Let us notice that we denote  $B_{r, s+1}^{-}=\left\{\lambda_j :  -1\leq\lambda_1\leq\cdots\leq\lambda_{r}\leq s-1\right\}$ in (\ref{Bk1}).

From (\ref{Bk}) it follows that 
\begin{eqnarray}
\fl
&&S^r_{s+1}(i)+T(i+r-2)T(i+s)\sum_{-1\leq\lambda_2\leq\cdots\leq\lambda_{r-1}\leq s}T(i+\lambda_2+r-2)\cdots T(i+\lambda_{r-1}+1)\nonumber \\
\fl
&&\;\;\;\;\;\;\;\;\;\;\;\;\;\;=S^r_{s}(i)+T(i+s)\sum_{\{\lambda_j\}\in B_{r, s+1}^{-}}T(i+\lambda_1+r-1)\cdots T(i+\lambda_{r-1}+1). \nonumber
\end{eqnarray}
Observe that the latter can be rewritten  as the identity
\begin{equation}
S^r_{s+1}(i)=Q^r_{s}(i)+T(i+s)S^{r-1}_{s+2}(i),
\label{s444}
\end{equation}
where, by definition, 
\begin{equation}
Q^r_s(i)=S^r_s(i)-T(i+r-2)T(i+s)S^{r-2}_{s+2}(i).
\label{s4}
\end{equation}
In a similar way we derive from (\ref{Bk1}) the identity 
\begin{equation}
S^r_{s+1}(i)=Q^r_s(i+1)+T(i+r-1)S^{r-1}_{s+2}(i) \label{s3} 
\end{equation}


Some remarks are in order. Relations (\ref{s5}) and (\ref{s4}) yield two mappings $\psi : \{T^r_s(i)\}\mapsto\{P^r_s(i)\}$ and $\varphi : \{S^r_s(i)\}\mapsto\{Q^r_s(i)\}$. It can be checked that $\psi\circ \varphi=\varphi\circ\psi$ and inverse mappings exist. It should be also noted that, in general, there is no problem to determine the $(h, n)$-classes of polynomials $\{P^r_s(i)\}$ and $\{Q^r_s(i)\}$  together with identities similar to (\ref{s444}), (\ref{s4}) and (\ref{s3}), but at the moment we do not see the need for this.

\subsection{Non-homogeneous polynomials}

Clearly, linear combinations (\ref{non}) and (\ref{non1}) in the case $(h, n)=(1, 1)$ become  
\begin{equation}
\tilde{T}^r_s(i)=\sum_{j=0}^{r}c_jT^{r-j}_{s+j}(i)
\label{nonhomT}
\end{equation}
 and
\begin{equation}
\tilde{S}^r_s(i)=
\left\{  
\begin{array}{l}
\displaystyle
\sum_{j=0}^{r}(-1)^{j}H_jS^{r-j}_{s-j}(i+j),\;\;\mbox{for}\;\; s>r  \\
\displaystyle
\sum_{j=0}^{s-1}(-1)^{j}H_jS^{r-j}_{s-j}(i+j),\;\;\mbox{for}\;\; s\leq r,
\end{array}
\right.
\label{nonhomS}
\end{equation}
 As for non-homogeneous polynomials $\tilde{P}^r_s(i)$ and $\tilde{Q}^r_s(i)$, they are defined by (\ref{nonhomT}) and (\ref{nonhomS}), respectively, with replacing $T\rightarrow P$ and $S\rightarrow Q$.

\subsection{The factorization property for polynomials $S^r_s(i)$}

Polynomials $S^r_s(i)$ have also the following property which will be used later on.
\begin{lem} \label{lem:2} 
Polynomials $S^r_s(i)$ satisfy  the factorization identity 
\begin{equation}
S^{r}_s(i)=S^{s-1}_{r+1}(i)\prod_{q=s-1}^{r-1}T(i+q) 
\label{sjk}
\end{equation}
for all $s\leq r$.
\end{lem}
 
Observe the following. For $s\leq r$, taking into account lemma \ref{lem:2}, we get
\begin{eqnarray}
\tilde{S}^r_s(i)&=&\sum_{j=0}^{s-1}(-1)^jH_jS^{r-j}_{s-j}(i+j)  \nonumber \\
&=&\left(\sum_{j=0}^{s-1}(-1)^jH_jS^{s-j-1}_{r-j+1}(i+j)\right)\prod_{q=s-1}^{r-1}T(i+q)  \nonumber \\ 
&=&\tilde{S}^{s-1}_{r+1}(i)\prod_{q=s-1}^{r-1}T(i+q).    \label{SKS}
\end{eqnarray}

\subsection{New binary operation on the sets of discrete polynomials} 

In addition to the operation $\left(\cdot, \cdot\right)$ given by  (\ref{ST21}) we introduce new one which will be needed in the following. Namely, let $\{F^r_s(i)\}$ and $\{G^r_s(i)\}$ be two arbitrary sets of discrete polynomials, such that $F^0_s(i)=G^0_s(i)\equiv 1$. Then, by definition,
\[
\langle F, G\rangle^r_s(i)= \sum_{j=0}^{r}(-1)^jF^{r-j}_{s+j}(i)G^j_{s-r+j}(i+r-j). 
\]
Remark that in this formula we put $F^r_s(i)\equiv 0$ for $s\leq 0$. For example,  $\langle F, G\rangle^r_1(i)=F^r_1(i)+(-1)^rG^r_1(i)$. Let
\[
\tilde{F}^r_s(i):= \sum_{j=0}^r(-1)^jH_jF^{r-j}_{s-j}(i+j)\;\;\mbox{and}\;\;\tilde{G}^r_s(i):=\sum_{j=0}^rc_jG^{r-j}_{s+j}(i).
\]
As can be checked by direct inspection, one has the following properties of this binary operation: 
\begin{equation}
\langle F, \tilde{G}\rangle^r_s(i)=\sum_{j=0}^{r-1}(-1)^jc_j\langle F, G\rangle^{r-j}_{s+j}(i)+(-1)^rc_r
\label{tildG}
\end{equation}
and
\begin{equation}
\langle\tilde{F}, G\rangle^r_s(i)=\sum_{j=0}^{r-1}(-1)^jH_j\langle F, G\rangle^{r-j}_{s-j}(i+j)+(-1)^rH_r. 
\label{tildF}
\end{equation}
We also observe that by virtue of (\ref{Tks})
\begin{equation}
\langle F, T \rangle^r_s(i) = F^r_s(i),\;\forall\; s\leq r-1
\label{identity8}
\end{equation}
for any set  of discrete polynomials $\{F^r_s(i)\}$. In the sequel, we use the following important conjecture. 
\begin{con} \label{lem:1}
The relations
\begin{eqnarray}
\langle S, T\rangle^r_s(i)&=&\sum_{j=0}^{r-2}(-1)^jT(i+r-j-2)T(i+s+j)\nonumber \\
&&\times S^{r-j-2}_{s+j+2}(i)T^j_{s-r+j}(i+r-j),\;\forall\; s\geq r-1
\label{identity1}
\end{eqnarray}
are identities. 
\end{con}
Unfortunately, by now we can prove this conjecture only for $r=1, 2, 3$. These proofs can be found in Appendix.

It is remarkable that by virtue of  (\ref{s5}) and (\ref{s4}), the identities of the form (\ref{identity1}) are equivalent to the following ones:
\begin{equation}
\langle S, P\rangle^r_s(i)\equiv 0,\;\forall\; s\geq r-1
\label{identity2}
\end{equation}
and
\begin{equation}
\langle Q, T\rangle^r_s(i)=0,\;\forall\; s\geq r-1.
\label{identity3}
\end{equation}
Moreover, from (\ref{tildG}), (\ref{tildF}), (\ref{identity2})  and (\ref{identity3}), it follows that  
\begin{equation}
\langle \tilde{S}, P\rangle^r_s(i)=(-1)^rH_r,\;\forall\; s\geq r-1
\label{identity6}
\end{equation}
and
\begin{equation}
\langle Q, \tilde{T}\rangle^r_s(i)=(-1)^rc_r,\;\forall\; s\geq r-1.
\label{identity7}
\end{equation}

\section{Ordinary difference equations in the case $(h, n)=(1, 1)$}
\label{sect4}

Clearly, in the case $(h, n)=(1, 1)$, equations of the form (\ref{12}) and (\ref{13})  are specified as
\begin{equation}
T(i+k+s)\tilde{T}^{k}_s(i)=T(i)\tilde{T}^{k}_s(i+2)
\label{122}
\end{equation}
and 
\begin{equation}
T(i+k)\tilde{S}^{k}_s(i)=T(i+s)\tilde{S}^{k}_s(i+2),
\label{132}
\end{equation}
respectively, while its equivalents (\ref{14}) and (\ref{15}) become 
\begin{equation}
T(i+k+s)\tilde{T}^{k}_{s-1}(i+1)=T(i)\tilde{T}^{k}_{s-1}(i+2)
\label{16}
\end{equation}
and 
\begin{equation}
T(i+k)\tilde{S}^{k}_{s+1}(i)=T(i+s)\tilde{S}^{k}_{s+1}(i+1),
\label{17}
\end{equation}
respectively. These equations are main objects to be studied throughout the rest of the paper.


We observe the following. Due to factorization property (\ref{SKS}), for $s\leq k$, equation  (\ref{132}) can be rewritten as 
\[
T(i+s-1)\tilde{S}^{s-1}_{k+1}(i)=T(i+k+1)\tilde{S}^{s-1}_{k+1}(i+2).
\]
This means that, for $s\leq k$, $(k, s)$th equation (\ref{132}) is equivalent to $(s-1, k+1)$th equation (\ref{132}) and therefore, without loss of generality, it is enough to consider this equation only for $s\geq k+1$.  

\subsection{Lax pairs and integrals} 
\label{subsec:41}

Here we discuss equations (\ref{122}) and (\ref{132}) from the point of view of their Lax pair representation \cite{Svinin6}.

\subsubsection{Lax pair for equation (\ref{132})}

A Lax pair for nonlinear difference equation (\ref{132}) is given by two linear equations: 
\begin{equation}
z\psi_{i+2}=z\psi_{i+1}+T(i+1)\psi_i
\label{3}
\end{equation}
and
\begin{equation}
w\psi_{i+k-s}=\sum_{j=0}^k(-1)^jz^{k-j}\tilde{S}^j_s(i+k-s-j+1)\psi_{i+k-j}
\label{44}
\end{equation}
which are a specification of (\ref{155}) and (\ref{157}) in the case $(h, n)=(1, 1)$. In what follows, we differ two cases: $s+k=2g+1$ and  $s+k=2g+2$, where $g\geq k$. 
Let  $\Psi_i:=\left(\psi^{(0)}_i,\ldots, \psi^{(s-1)}_i\right)^{T}$. One sees that a pair of linear equations (\ref{3}) and (\ref{44}) is equivalent to two linear systems
\begin{equation}
L_i\Psi_i=0,\;\;
\Psi_{i+1}=A_i\Psi_i
\label{LA}
\end{equation}
the first one of which is given exactly by 
\[
\begin{array}{l}
T(i)\psi_{i-1}+z\psi^{(0)}_i-z\psi^{(1)}_i=0, \\ [0.3cm]
T(i+1)\psi^{(0)}_i+z\psi^{(1)}_i-z\psi^{(2)}_i=0,\ldots \\ [0.3cm]
T(i+s-2)\psi^{(s-3)}_i+z\psi^{(s-2)}_i-z\psi^{(s-1)}_i=0, \\ [0.3cm]
T(i+s-1)\psi^{(s-2)}_i+z\psi^{(s-1)}_i-z\psi_{i+s}=0,
\end{array}
\]
while the second one is given by
\[
\psi^{(0)}_{i+1}=\psi^{(1)}_{i},\ldots, \psi^{(s-2)}_{i+1}=\psi^{(s-1)}_{i}, \psi^{(s-1)}_{i+1}=\psi_{i+s},
\]
where
\[
\psi_{i-1}=\frac{1}{w}\sum_{j=0}^k(-1)^jz^{k-j}\tilde{S}^j_s(i-j)\psi^{(s-j-1)}_i
\]
and
\[
\psi_{i+s}=\frac{w}{z^k}\psi^{(0)}_i-\sum_{j=1}^k(-1)^jz^{-j}\tilde{S}^j_s(i-j+1)\psi^{(s-j)}_i.
\]
Checking the condition $\det(L)= 0$ give the following. In the case $k+s=2g+1$, we get
\begin{equation}
z^{2g-2k+1}w^2-z^{g-k+1}R_g(z)w-\mathcal{H}_0\tilde{R}_k(z)=0,
\label{spectr1}
\end{equation}
while in the case $k+s=2g+2$, we obtain
\begin{equation}
z^{2g-2k+2}w^2-z^{g-k+1}R_{g+1}(z)w+\mathcal{H}_0\tilde{R}_k(z)=0, 
\label{spectr2}
\end{equation}
where
\begin{equation}
R_g(z)=z^g+\sum_{j=1}^gH_jz^{g-j},\;\;
\tilde{R}_k(z)=z^k+\sum_{j=1}^k\tilde{c}_jz^{k-j}
\label{Rg}
\end{equation}
and $\tilde{c}_r\equiv \sum_{q=0}^{r}c_qH_{r-q}$. Here $\{H_1,\ldots, H_k\}$ is the set of parameters entering in equation (\ref{132}), while  $\mathcal{H}_0^{(k, s)}\equiv\mathcal{H}_0$ and $c_r^{(k, s)}\equiv c_r$ for $r=1,\ldots, k$ and $H_r^{(k, s)}\equiv H_r$, for $r\geq k+1$,  are supposed to be nontrivial first integrals of $(k, s)$th equation (\ref{132}). In particular,
\[
\mathcal{H}_0^{(k, s)}(i)=(-1)^k\tilde{S}^k_{s+1}(i)\prod_{q=k}^{s-1}T(i+q).
\]
In section \ref{sect5}, we describe these integrals regardless of Lax representation (\ref{LA}).

\subsubsection{Lax pair for equation (\ref{122})} 

Now let us consider a Lax pair completed by linear equations (\ref{3}) and  
\begin{equation}
w\psi_{i+k+s}=\sum_{j=0}^kz^{k-j}\tilde{T}^j_s(i+k-j+1)\psi_{i+k-j}
\label{4}
\end{equation}
which is a specification of (\ref{156}). 
Let  $\Psi_i:=\biggl(\psi^{(0)}_i,\ldots, \psi^{(k+s-1)}_i\biggr)^{T}$. We rewrite two linear equations 
(\ref{3}) and (\ref{4}) in the form (\ref{LA}) or more exactly as a system
\[
\begin{array}{l}
T(i)\psi_{i-1}+z\psi^{(0)}_i-z\psi^{(1)}_i=0, \\ [0.3cm]
T(i+1)\psi^{(0)}_i+z\psi^{(1)}_i-z\psi^{(2)}_i=0,\ldots \\ [0.3cm]
T(i+k+s-2)\psi^{(k+s-3)}_i+z\psi^{(k+s-2)}_i-z\psi^{(k+s-1)}_i=0, \\ [0.3cm]
T(i+k+s-1)\psi^{(k+s-2)}_i+z\psi^{(k+s-1)}_i-z\psi_{i+k+s}=0,
\end{array}
\]
and
\[
\psi^{(0)}_{i+1}=\psi^{(1)}_{i},\ldots, \psi^{(k+s-2)}_{i+1}=\psi^{(k+s-1)}_{i}, \psi^{(k+s-1)}_{i+1}=\psi_{i+k+s}
\]
with
\[
\psi_{i-1}=\frac{1}{\tilde{T}^k_s(i)}\biggl( w\psi^{(k+s-1)}_{i}-\sum_{j=0}^{k-1}z^{k-j}\tilde{T}^j_s(i+k-j)\psi^{(k-j-1)}_{i} \biggr)
\]
and
\[
\psi_{i+k+s}=\frac{1}{w}\sum_{j=0}^kz^{k-j}\tilde{T}^j_s(i+k-j+1)\psi^{(k-j)}_{i}.
\]
Actual calculations give the following. In the case $k+s=2g+1$, we get 
\begin{equation}
\mathcal{H}_0w^2+z^{g+1}R_g(z)w-z^{2g+1}\tilde{R}_k(z)=0,
\label{T,2g+1}
\end{equation}
while in the case $k+s=2g+2$, we obtain
\begin{equation}
\mathcal{H}_0w^2-z^{g+1}R_{g+1}(z)w+z^{2g+2}\tilde{R}_k(z)=0, 
\label{T,2g+2}
\end{equation}
where $R_g(z)$ and $\tilde{R}_k(z)$  are as in (\ref{Rg}), but  $\{\mathcal{H}_0, H_1,\ldots, H_g\}$  in this case is the set of the first integrals depending on  $c_j$ being the parameters contained in difference equation (\ref{122}). The curves given by equations (\ref{T,2g+1}) and (\ref{T,2g+2}) are hyperelliptic ones. Indeed, applying  birational transformation 
\[ 
z=\frac{1}{Z},\;\;\;
w=\frac{W-r_g(Z)}{2\mathcal{H}_0Z^{2g+1}}
\] 
to equation (\ref{T,2g+1}), we obtain that it becomes
\[
W^2=r_g^2(Z)+4\mathcal{H}_0Z^{2g-k+1}\tilde{r}_k(Z),
\]
while transforming  (\ref{T,2g+2}) with the help of
\[
z=Z,\;\;\;
w=Z^{g+1}\frac{W+R_{g+1}(Z)}{2\mathcal{H}_0}
\]
yields
\[
W^2=R_{g+1}^2(Z)-4\mathcal{H}_0\tilde{R}_k(Z).
\]
Also, it is important to notice that using the transformation 
\[
w\mapsto \frac{z^{g+1}}{\mathcal{H}_0}\left(z^{g-k}w-R_g(z)\right)
\]
for (\ref{T,2g+1}) gives (\ref{spectr1}), while the transformation
\[
w\mapsto \frac{z^{g+1}}{\mathcal{H}_0}\left(R_{g+1}(z)-z^{g-k+1}w\right)
\]
converts (\ref{T,2g+2}) to (\ref{spectr2}). This suggests that associated difference  equations have to be closely related to each other and we discuss this question in the following.   

\section{First integrals}
\label{sect5}

\subsection{Integrals for equation (\ref{132})}

Let us first consider an infinite class of equations of the form (\ref{132}) for $s\geq k+1$. We denote the solution space of (\ref{132}) by $\mathcal{N}^k_s$. Every solution of this equation is defined by a set of initial data $\left\{y_j=T(i_0+j) : j=0,\ldots, k+s-1\right\}$ and constants $\{H_1,\ldots, H_k\}$. Thus, the dimension of $\mathcal{N}^k_s$ is $2k+s$. 


From equivalent form of equation (\ref{132}) given by (\ref{17}), it is evident that the polynomial\footnote{Equations of the form $G_0=\tilde{S}^1_{s+1}(i)\prod_{q=1}^{s-1}T(i+q)$ with arbitrary constant $G_0$ and $H_1=0$ has appeared and studied in \cite{Demskoi}.}  
\[
G_0^{(k, s)}(i)=\tilde{S}^k_{s+1}(i)\prod_{q=k}^{s-1}T(i+q) 
\]
is its first integral.  To construct  some number of integrals for this equation, let us consider   discrete polynomials of the form
\begin{equation}
G_r^{(k, s)}(i)=\sum_{j=0}^rQ^j_{s-k-j-1}(i+k+1)G_0^{(k+j, s-j)}(i).
\label{int}
\end{equation}
Let us remember that $Q^r_s(i)$ are the discrete polynomials defined by (\ref{s4}). Clearly, these polynomials satisfy recurrent relations 
\begin{equation}
G_{r+1}^{(k, s)}(i)=G_r^{(k, s)}(i)+Q^{r+1}_{s-k-r-2}(i+k+1)G_0^{(k+r+1, s-r-1)}(i).
\label{rec}
\end{equation}
\begin{pro}    \label{lemm1}
The polynomial (\ref{int}), for any fixed $k, r\geq 0$ and $s\geq k+2r+1$ is the first integral of $(k+r, s-r)$th equation (\ref{132}). Moreover the relation
\begin{eqnarray}
G_r^{(k, s)}(i+1)-G_r^{(k, s)}(i)&=&\Lambda_{r}^{(k, s)}(i)\left(T(i+s-r)\tilde{S}^{k+r}_{s-r+1}(i+1) \right.\nonumber \\
&&\left. -T(i+k+r)\tilde{S}^{k+r}_{s-r+1}(i)\right)  \label{l1} \\
&=&\Lambda_{r}^{(k, s)}(i)\left(\tilde{S}^{k+r+1}_{s-r}(i+1)-\tilde{S}^{k+r+1}_{s-r}(i)\right), \label{l2}  
\end{eqnarray}
with
\[
\Lambda_{r}^{(k, s)}(i)\equiv S^r_{s-k-r}(i+k+1)\prod_{q=k+r+1}^{s-r-1}T(i+q)
\]
being a corresponding integrating factor, is valid.
\end{pro}
Therefore $\{\mathcal{G}_r^{(k, s)}(i)\equiv G_r^{(k-r, s+r)}(i) : r=0,\ldots, k\}$ is a multitude of the first integrals for  $(k, s)$th equation (\ref{132}). 
We easily deduce the system 
\begin{equation}
\mathcal{G}^{(k, s)}(i)=Q_{s-k}(i+k)G^{(k, s)}(i)
\label{Gks}
\end{equation} 
with
\[
\mathcal{G}^{(k, s)}(i)=\left(\mathcal{G}_0^{(k, s)}(i), \mathcal{G}_1^{(k, s)}(i),\ldots, \mathcal{G}_k^{(k, s)}(i)\right)^{T}
\] 
and  
\[
G^{(k, s)}(i)=\left(G_0^{(k, s)}(i), G_0^{(k-1, s+1)}(i),\ldots, G_0^{(0, s+k)}(i)\right)^{T}
\] 
which follows from (\ref{int}). A matrix $Q_s(i)$ in (\ref{Gks}) is supposed to be $(k+1)\times (k+1)$ matrix with  elements $(Q_s(i))_{pq}=Q^{p-q}_{s+p+q-3}(i-p+2)$.


From (\ref{l1}) and (\ref{l2}), it is obvious that
\begin{eqnarray}
\fl
\mathcal{G}_r^{(k, s)}(i+1)-\mathcal{G}_r^{(k, s)}(i)&=&\Lambda_{r}^{(k-r, s+r)}(i)\left(T(i+s)\tilde{S}^{k}_{s+1}(i+1)-T(i+k)\tilde{S}^{k}_{s+1}(i)\right)
   \nonumber \\
\fl
&=&\Lambda_{r}^{(k-r, s+r)}(i)\left(\tilde{S}^{k+1}_{s}(i+1)-\tilde{S}^{k+1}_{s}(i)\right). \nonumber  
\end{eqnarray}


In what follows, it is useful also to consider the following linear combinations of integrals: 
\begin{equation}
\tilde{\mathcal{G}}_r^{(k, s)}(i)\equiv  \sum_{q=0}^r(-1)^qH_q\mathcal{G}_{r-q}^{(k, s)}(i),\;\;
r=0,\ldots, k. 
\end{equation}
One can see that $\tilde{\mathcal{G}}_r^{(k, s)}(i)=\tilde{G}_r^{(k-r, s+r)}(i)$ with
\[
\tilde{G}_r^{(k, s)}(i)\equiv\sum_{j=0}^r\tilde{Q}^j_{s-k-j-1}(i+k+1)G_0^{(k+j, s-j)}(i)
\]
and
\begin{eqnarray}
\tilde{\mathcal{G}}_r^{(k, s)}(i+1)-\tilde{\mathcal{G}}_r^{(k, s)}(i)&=&\tilde{\Lambda}_{r}^{(k, s)}(i)\left(T(i+s)\tilde{S}^{k}_{s+1}(i+1)-T(i+k)\tilde{S}^{k}_{s+1}(i)\right)
   \nonumber \\
&=&\tilde{\Lambda}_{r}^{(k, s)}(i)\left(\tilde{S}^{k+1}_{s}(i+1)-\tilde{S}^{k+1}_{s}(i)\right) \nonumber  
\end{eqnarray}
with $\tilde{\Lambda}_{r}^{(k, s)}(i)=G_0^{(r, s-k+r-1)}(i+k-r+1)$. 


Let us consider
\begin{equation}
\tilde{G}_r^{(0, s)}(i)=\sum_{j=0}^r\tilde{Q}^j_{s-j-1}(i+1)\tilde{S}^{j}_{s-j+1}(i)\prod_{q=j}^{s-j-1} T(i+q).
\label{linearcomb0}
\end{equation}
Using the identity of the form (\ref{s1}), namely, 
\[
\tilde{S}^{j}_{s-j+1}(i)=\tilde{S}^{j}_{s-j}(i+1)+T(i+j-1)\tilde{S}^{j-1}_{s-j+1}(i),
\]
let us present (\ref{linearcomb0}) as
\begin{eqnarray}
\tilde{G}_r^{(0, s)}(i)&=&\sum_{j=0}^rg_j^s(i)+\sum_{j=1}^rf_j^s(i)\nonumber  
\end{eqnarray}
with
\[
g_j^s(i)\equiv \tilde{Q}^j_{s-j-1}(i+1)\tilde{S}^{j}_{s-j}(i+1)\prod_{q=j}^{s-j-1} T(i+q)
\]
and
\[
f_j^s(i)\equiv \tilde{Q}^j_{s-j-1}(i+1)\tilde{S}^{j-1}_{s-j+1}(i)\prod_{q=j-1}^{s-j-1} T(i+q).
\]
In the sequel, we need the following lemma.
\begin{lem} \label{lemm:2}
The relations:
\begin{equation}
\sum_{j=0}^rg_{j}^s(i)+\sum_{j=1}^{r+1}f_{j}^s(i)=\tilde{S}^r_{s-r}(i)\tilde{S}^{r+1}_{s-r-1}(i+1)\prod_{q=r}^{s-r-2}T(i+j)
\label{1111}
\end{equation}
and
\begin{equation}
\sum_{q=0}^rg_{j}^s(i)+\sum_{q=1}^rf_{j}^s(i)=\tilde{S}^r_{s-r}(i)\tilde{S}^{r}_{s-r}(i+1)\prod_{q=r}^{s-r-1}T(i+q)
\label{2222}
\end{equation}
are identities.
\end{lem}
Therefore by virtue of this lemma, 
\begin{equation}
\tilde{\mathcal{G}}_k^{(k, s)}=\tilde{G}_k^{(0, s+k)}(i)=\tilde{S}^k_{s}(i)\tilde{S}^{k}_{s}(i+1)\prod_{q=k}^{s-1}T(i+q).
\label{3333}
\end{equation}
Here we have replaced $r\rightarrow k$ and $s-r\rightarrow s$. It should be noted that we got in fact an expected result since (\ref{3333}) is an obvious first integral of $(k, s)$th equation  (\ref{132}). 

\subsection{Chains of inclusions of $\mathcal{N}^k_s$}

Suppose that, for some $k\geq 1$ and $s\geq k+1$, the condition 
\begin{equation}
\tilde{S}^k_{s+1}(i)\equiv 0
\label{co1}
\end{equation}
is fulfilled. It can be easily seen that 
\[
\mathcal{G}_r^{(k, s)}|_{(\ref{co1})}=\mathcal{G}_{r-1}^{(k-1, s+1)},\;\;\;
r=1,\ldots, k.
\] 
This means that under this condition the set of integrals $\{\mathcal{G}_r^{(k, s)} : r=0,\ldots, k\}$ for $(k, s)$th equation (\ref{17}) is reduced to the set $\{\mathcal{G}_r^{(k-1, s+1)} : r=0,\ldots, k-1\}$. This analysis of the first integrals suggests that it is possible that $\mathcal{N}^{k-1}_{s+1}\subset \mathcal{N}^{k}_{s}$. To clarify this question, let us first consider the simplest case $k=1$. One can rewrite $(1, s)$th equation (\ref{17})  as 
\begin{equation}
T(i+s+1)=\frac{T(i+1)}{T(i+s)}\left(\sum_{j=1}^{s+1}T(i+j-1)-H_1\right)-\sum_{j=1}^{s}T(i+j)+H_1.
\label{1s}
\end{equation}
Suppose now that 
\begin{equation}
\tilde{S}^1_{s+1}(i)\equiv \sum_{j=1}^{s+1}T(i+j-1)-H_1=0.
\label{constr}
\end{equation}
By direct inspection, one can check that this constraint is compatible with (\ref{1s}). Indeed, by virtue of (\ref{1s}), one has
\[
\tilde{S}^1_{s+1}(i+1)=\frac{T(i+1)}{T(i+s)}\tilde{S}^1_{s+1}(i).
\]
In turn, equation (\ref{constr}) is equivalent to $(0, s+1)$th equation (\ref{17}), which is the periodicity one: $T(i+s+1)=T(i)$. 


Now let us turn to the general situation. Rewriting   $(k, s)$th equation (\ref{17})  as
\begin{equation}
\tilde{S}^k_{s+1}(i+1)=\frac{T(i+k)}{T(i+s)}\tilde{S}^k_{s+1}(i)=0
\label{25}
\end{equation}
we immediately see  that condition (\ref{co1}) is evidently compatible with this equation. Moreover we observe that, taking into account  identities  (\ref{s1}) and (\ref{s2}), we can rewrite this restriction as $(k-1, s+1)$th equation (\ref{17}), that is, 
\begin{equation}
T(i+k-1)\tilde{S}^{k-1}_{s+2}(i)=T(i+s+1)\tilde{S}^{k-1}_{s+2}(i+1).
\label{sk}
\end{equation}
For any solution of this equation corresponding to some fixed initial data $\left\{y_j : j=0,\ldots, k+s-1\right\}$ and  a set  of the parameters $\{H_1,\ldots, H_{k-1}\}$ we calculate the constant $H_k$ through the condition (\ref{co1}), where one substitute $T(i+j)=y_j$ for $j=0,\ldots, k+s-1$.  For example, in the case $k=1$ we have to put $H_1=\sum_{j=0}^{s}y_{j}$.
Then we can assert that this solution of (\ref{sk}) also solves $(k, s)$th equation (\ref{17}) with the same initial data $\left\{y_j : j=0,\ldots, k+s-1\right\}$ and  a set  of the parameters $\{H_1,\ldots, H_{k}\}$. Therefore, we conclude that $\mathcal{N}^{k-1}_{s+1}\subset\mathcal{N}^k_s$.


Given some $g\geq k$, there are two different cases to consider: $s+k=2g+1$ and $s+k=2g+2$. For these two cases we have two chains of inclusions
\[
\mathcal{N}_{2g+1}^0\subset\mathcal{N}_{2g}^1\subset \cdots\subset\mathcal{N}_{g+1}^g
\]
and
\[
\mathcal{N}_{2g+2}^0\subset\mathcal{N}_{2g+1}^1\subset \cdots\subset\mathcal{N}_{g+2}^g,
\]
respectively. 


Consider now the solution space $\mathcal{N}_{g+2}^g$ corresponding to $(g, g+2)$th equation (\ref{132}), that is,
\[
T(i+g)\tilde{S}^g_{g+2}(i)=T(i+g+2)\tilde{S}^g_{g+2}(i+2)
\]  
and next observe that this equation is a consequence of the relation
\begin{equation}
T(i+g)\tilde{S}^g_{g+2}(i)=T(i+g+1)\tilde{S}^g_{g+2}(i+1)
\label{gg}
\end{equation}
which is $(g, g+1)$th equation (\ref{17}).  This means that $\mathcal{N}^{g}_{g+1}\subset\mathcal{N}^{g}_{g+2}$. 


Now we are in position to formulate the following theorem. 
\begin{thm}
All the solution spaces $\mathcal{N}^k_s$ are organized in the following diagram of inclusions:
\[
\fl
\begin{array}{ccccccccccccccc}
{\mathcal N}_1^0 & \subset & {\mathcal N}_2^0 &  {\mathcal N}_3^0 &          & {\mathcal N}_4^0 & &{\mathcal N}_5^0 &         & {\mathcal N}_6^0 & & {\mathcal N}_7^0 & & {\mathcal N}_8^0 & \cdots  \\
             & & &       \cap    &          &    \cap      & &   \cap      &  &  \cap        & &  \cap       &  & \cap                      \\ 
             & & &  {\mathcal N}_2^1 & \subset  & {\mathcal N}_3^1 & & {\mathcal N}_4^1 & & {\mathcal N}_5^1 & & {\mathcal N}_6^1 &  & {\mathcal N}_7^1 & \cdots  \\
             & & &               &          &              & &   \cap      &  &  \cap        &  & \cap       & & \cap &                     \\
             & & &               &          &              & & {\mathcal N}_3^2 & \subset & {\mathcal N}_4^2 & & {\mathcal N}_5^2 & & {\mathcal N}_6^2 & \cdots  \\
             & & &               &          &              & &              & &  &  & \cap            &              &  \cap        &  \\
             & & &               &          &              & &              & &  &  & {\mathcal N}_4^3    & \subset      & {\mathcal N}_5^3 &  \cdots
\end{array}
\]
\end{thm}
Let us remark that from this theorem it follows  that every equation (\ref{132})  admits  periodic solutions. 


Let us consider now the set of integrals $\{\mathcal{G}_j^{(g, g+2)} : j=1, \ldots, g\}$ and investigate how they behave when restricting them on $\mathcal{N}^g_{g+1}$. To this aim, it is useful the following technical lemma.
\begin{lem} \label{lemm:1}
The relations
\begin{eqnarray}
G_r^{(k, s)}(i)-G_{r+1}^{(k-1, s)}(i)&=&S^r_{s-k-r}(i+k)\tilde{S}^{k+r}_{s-r}(i+1)\prod_{q=k+r}^{s-r-1}T(i+q) \nonumber \\
&&-S^{r+1}_{s-k-r-1}(i+k)\tilde{S}^{k+r}_{s-r}(i)\prod_{q=k+r}^{s-r-2}T(i+q) \label{lemm}
\end{eqnarray}
are identities. 
\end{lem}
Putting $k=g-r$ and $s=g+r+2$ into (\ref{lemm}) gives 
\begin{eqnarray}
\mathcal{G}_r^{(g, g+2)}(i)-\mathcal{G}_{r+1}^{(g, g+1)}(i)&\equiv &G_r^{(g-r, g+r+2)}(i)-G_{r+1}^{(g-r-1, g+r+2)}(i)\nonumber \\
&=&S^r_{r+2}(i+g-r)\tilde{S}^g_{g+2}(i+1)T(i+g)T(i+g+1) \nonumber \\
&&-S^{r+1}_{r+1}(i+g-r)\tilde{S}^g_{g+2}(i)T(i+g). \nonumber
\end{eqnarray}
Therefore, we obtain 
\[
\fl
\mathcal{G}_r^{(g, g+2)}(i)|_{(\ref{gg})}-\mathcal{G}_{r+1}^{(g, g+1)}(i)=T(i+g)\tilde{S}^g_{g+2}(i)\left(T(i+g)S^r_{r+2}(i+g-r)-S^{r+1}_{r+1}(i+g-r)\right). 
\]
In turn, by virtue of the factorization property (\ref{sjk}), the latter is identically zero. Therefore we can conclude that when restricting $\mathcal{N}^g_{g+2}$ on $\mathcal{N}^g_{g+1}$ one has 
\[
\mathcal{G}_r^{(g, g+2)}|_{(\ref{gg})}=\mathcal{G}_{r+1}^{(g, g+1)}\;\;\;
r=0,\ldots, g-1.
\] 
Let us remark that  this set of relations can be also written as
\[
\tilde{\mathcal{G}}_r^{(g, g+2)}|_{(\ref{gg})}\equiv \tilde{\mathcal{G}}_{r+1}^{(g, g+1)}+(-1)^rH_{r+1}G_0^{(g, g+1)}. 
\] 


Finally, we want to find out what happens with the integral $\mathcal{G}_g^{(g, g+2)}$ under this restriction. It turns out more convenient to consider $\tilde{\mathcal{G}}_g^{(g, g+2)}$ instead of $\mathcal{G}_g^{(g, g+2)}$. 
Putting $k=g$ and $s=g+2$ into (\ref{3333}) yields the identity
\[
\tilde{\mathcal{G}}_g^{(g, g+2)}=\tilde{S}^g_{g+2}(i)\tilde{S}^{g}_{g+2}(i+1)T(i+g)T(i+g+1).
\]
Therefore we get that  restricting of the integral $\tilde{\mathcal{G}}_g^{(g, g+2)}$ on $\mathcal{N}^g_{g+1}$ yields
\[
\tilde{\mathcal{G}}_g^{(g, g+2)}(i)|_{(\ref{gg})}=\left(\tilde{S}^g_{g+2}(i)T(i+g)\right)^2=\left(G_0^{(g, g+1)}(i)\right)^2.
\]

\subsection{Equivalence of equations (\ref{122}) and (\ref{132})}  
\label{sect5.3}

Consider a set of the first integrals  
\begin{eqnarray}
c_r^{(k, s)}(i)&=&(-1)^r\frac{\mathcal{G}_r^{(k, s)}(i)}{G_0^{(k, s)}(i)} \nonumber \\ 
 							&=&(-1)^r\sum_{j=0}^rQ^j_{s-k+2r-j-1}(i+k-r+1)\xi_{r-j}^{(k, s)}(i),\;\; r=1,\ldots, k	 \label{crks} 
\end{eqnarray}
for equation (\ref{132}). Here 
\begin{eqnarray}
\xi_r^{(k, s)}(i)&\equiv &\frac{G_0^{(k-r, s+r)}(i)}{G_0^{(k, s)}(i)}\nonumber \\ 
                 &= &\frac{\tilde{S}^{k-r}_{s+r+1}(i)}{\tilde{S}^k_{s+1}(i)}\prod_{q=1}^rT(i+k-q)\prod_{q=1}^rT(i+s+q-1). \nonumber 
\end{eqnarray}
Let $c_r^{(k, s)}$ be some values of the integrals given by (\ref{crks}) corresponding to some initial data $\{y_0,\ldots, y_{s+k-1}\}$ and set of parameters $\{H_1,\ldots, H_k\}$.  In what follows it makes sense to use  simplified notation $c_r^{(k, s)}\equiv c_r$. 
The  system equations
\begin{equation}
c_r=(-1)^r\sum_{j=0}^{r}Q^j_{s-k+2r-j-1}(i+k-r+1)\xi_{r-j}^{(k, s)}(i),\;\;\; 
r=1,\ldots, k
\label{cH}
\end{equation}
yields some relationship between the set of parameters $\{H_r\}$ of $(k, s)$th equation (\ref{17}) and its integrals $\{c_r\}$.  


The following question arises: what would happen if we interchange $\{H_r\}$ and $\{c_r\}$? We claim the following: in a result, we obtain $(k, s)$th equation (\ref{16}). Before discussing the general situation, let us first illustrate  this  on simple example  $k=1$.\footnote{It is obvious that in the case $k=0$ both the equations (\ref{16}) and (\ref{17})   are periodicity one: $T(i+s)=T(i)$.} In this case, system (\ref{cH}) is specified as a single equation
\begin{equation}
c_1=-Q^1_{s-1}(i+1)-\frac{T(i)T(i+s)}{\tilde{S}^1_{s+1}(i)}. 
\label{c1}
\end{equation}
Since $Q^1_{s-1}(i)=T^1_{s-1}(i)\equiv\sum_{j=1}^{s-1}T(i+j-1)$, then we can rewrite (\ref{c1}) as
\[
\tilde{S}^1_{s+1}(i)=-\frac{T(i)T(i+s)}{\tilde{T}^1_{s-1}(i+1)}\;\;\;\mbox{or}\;\;\; 
\xi_1^{(1, s)}(i)=-\tilde{T}^1_{s-1}(i+1).
\]
One can check that substituting the latter into $(1, s)$th equation (\ref{17}) gives  $(1, s)$th equation (\ref{16}). Therefore we proved that in the case $k=1$, equations (\ref{132}) and (\ref{122}), for any fixed $s\geq 2$, yield the same dynamical system. Given some value of $H_1$ and initial data $\left\{y_0,\ldots, y_{s}\right\}$ we calculate the corresponding parameter $c_1$ with the help of
\[
\displaystyle
c_1=-\sum_{j=1}^{s-1}y_j-\frac{y_0y_{s}}{\sum_{j=0}^{s}y_j-H_1}.
\] 
In turn, resolving the latter in favor of $H_1$ gives
\[
\displaystyle
H_1=\sum_{j=0}^{s}y_j+\frac{y_0y_{s}}{\sum_{j=1}^{s-1}y_j+c_1}.
\] 
Therefore in this case $c_1$ and $H_1$ are related to each other by birational transformation. Clearly,  requiring  $H_1=\sum_{j=0}^{s}y_j$ yields $c_1=\infty.$


Now let us turn to the general situation. We would like to show that system of equations (\ref{cH}) is equivalent to the following one:
\begin{equation}
\xi_r^{(k, s)}(i)=(-1)^r\tilde{T}^r_{s-k+r-1}(i+k-r+1),\;\;r=1,\ldots, k.
\label{xij}
\end{equation}
Substituting (\ref{xij}) into (\ref{cH}) gives
\begin{eqnarray}
c_r&=&\sum_{j=0}^{r}(-1)^{j}Q^j_{s-k+2r-j-1}(i+k-r+1)\tilde{T}^{r-j}_{s-k+r-j-1}(i+k-r+j+1) \nonumber \\
&\equiv&(-1)^r\langle Q, \tilde{T}\rangle^r_{s-k+r-1}(i+k-r+1).  \nonumber
\end{eqnarray}
The latter is an identity provided that (\ref{identity7}) does.


Let us write down the last equation in system (\ref{xij}) as
\begin{equation}
(-1)^k\tilde{T}^k_{s-1}(i+1)\tilde{S}^k_{s+1}(i)=\prod_{q=1}^kT(i+k-q)\prod_{q=1}^kT(i+s+q-1)
\label{xik}
\end{equation}
and observe that replacing in  (\ref{17}) the polynomial $\tilde{S}^k_{s+1}(i)$  by $\tilde{T}^k_{s-1}(i)$ by virtue of (\ref{xik}) yields (\ref{16}). Therefore we prove the equivalence of two  discrete equations (\ref{16}) and  (\ref{17}) or their equivalents  (\ref{122}) and (\ref{132})   for any fixed $k\geq 0$ and $s\geq k+1$ and show how parameters $c_r$ and $H_r$ are related to each other. Of course, it is valid provided that conjecture \ref{lem:1} does.


Finally, we would like to resolve  system  (\ref{xij}) in favor of $H_r$. To this aim, we rewrite this system as
\begin{eqnarray}
&&(-1)^r\tilde{T}^r_{s-k+r-1}(i+k-r+1)\tilde{S}^k_{s+1}(i)\nonumber\\
&&\;\;\;\;\;\;\;\;\;=\tilde{S}^{k-r}_{s+r+1}(i)\prod_{q=1}^rT(i+k-q)\prod_{q=1}^rT(i+s+q-1),\;\;
r=1,\ldots, k\nonumber
\end{eqnarray}
and then observe that it is equivalent to the following system:
\begin{eqnarray}
&&(-1)^r\tilde{S}^r_{s+k-r+1}(i)\tilde{T}^k_{s-1}(i+1)\nonumber\\
&&\;\;\;\;\;\;\;\;\;=\tilde{T}^{k-r}_{s-r-1}(i)\prod_{q=1}^rT(i+q-1)\prod_{q=1}^rT(i+s+k-q),\;\;
r=1,\ldots, k\nonumber
\end{eqnarray}
which in turn we may rewrite as
\begin{equation}
\eta_r^{(k, s)}(i)=(-1)^r\tilde{S}^r_{s+k-r+1}(i),\;\;
r=1,\ldots, k
\label{eta}
\end{equation}
with
\begin{equation}
\eta_r^{(k, s)}(i)\equiv\frac{\tilde{T}^{k-r}_{s-r-1}(i+r+1)}{\tilde{T}^k_{s-1}(i+1)}\prod_{q=1}^rT(i+q-1)\prod_{q=1}^rT(i+s+k-q).
\label{eta1}
\end{equation}
We can solve (\ref{eta}) as
\begin{equation}
H_r=\sum_{j=0}^{r}P^j_{s+k-2r+j+1}(i+r-j)\eta_{r-j}^{(k, s)}(i).
\label{Hj}
\end{equation}
 Indeed, substituting (\ref{eta}) into (\ref{Hj}) gives 
$
H_r=(-1)^r\langle\tilde{S}, P\rangle^r_{s+k-r+1}(i).  
$
By virtue of (\ref{identity6}), the latter is an identity. Note that relation (\ref{Hj}) gives  some number of integrals for $(k, s)$th equation (\ref{16}).  

\subsection{Integrals for  equation (\ref{16})}

Inspired by (\ref{Hj}) we consider the rational discrete function 
\begin{equation}
F_r^{(k, s)}(i)=\sum_{j=0}^rP^j_{s+k+j+1}(i-j)F_0^{(k+j, s+j)}(i-j),
\label{int1}
\end{equation}
where
\[
F_0^{(k, s)}(i)\equiv \frac{\tilde{T}^k_{s-1}(i+1)}{\prod_{q=0}^{s+k-1}T(i+q)}
\]
is obvious integral of $(k, s)$th equation (\ref{16}). It is important to notice that $k$ in (\ref{int1}) is allowed to take negative values. In this case we put $F_0^{(r,s )}(i)\equiv 0,\; \forall r<0$. Clearly, this rational function satisfies recurrent relation
\begin{equation}
F_{r+1}^{(k, s)}(i)=F_r^{(k, s)}(i)+P^{r+1}_{s+k+r+2}(i-r-1)F_0^{(k+r+1, s+r+1)}(i-r-1).
\label{rec1}
\end{equation}
\begin{pro}    \label{pro:3}
The rational function (\ref{int1}) is a first integral of $(k+r, s+r)$th equation (\ref{16}). Moreover the relation
\begin{eqnarray}
\fl
F_r^{(k, s)}(i+1)-F_r^{(k, s)}(i)&=&\Delta_{r}^{(k, s)}(i)\left(T(i-r)\tilde{T}^{k+r}_{s+r-1}(i-r+2) \right.\nonumber \\
\fl
&&\left. -T(i+k+s+r)\tilde{T}^{k+r}_{s+r-1}(i-r+1)\right)  \label{l11} \\
\fl
&=&\Delta_{r}^{(k, s)}(i)\left(\tilde{T}^{k+r+1}_{s+r}(i-r)-\tilde{T}^{k+r+1}_{s+r}(i-r+1)\right), \label{l21}  
\end{eqnarray}
with an integrating factor
\[
\Delta_{r}^{(k, s)}(i)\equiv \frac{T^r_{s+k+r}(i-r+1)}{\prod_{q=-r}^{s+k+r}T(i+q)}
\]
is valid.
\end{pro}

Therefore $\{\mathcal{F}_r^{(k, s)}(i)\equiv F_r^{(k-r, s-r)}(i+r) \}$ presents a multitude of the first integrals for  $(k, s)$th equation (\ref{16}).  
Moreover, from (\ref{l11}) and (\ref{l21}), it follows that
\begin{eqnarray}
\fl
\mathcal{F}_r^{(k, s)}(i+1)-\mathcal{F}_r^{(k, s)}(i)&=&\Delta_{r}^{(k-r, s-r)}(i+r)\left(T(i)\tilde{T}^{k}_{s-1}(i+2)-T(i+k+s)\tilde{T}^{k}_{s-1}(i+1)\right)
   \nonumber \\
\fl
&=&\Delta_{r}^{(k-r, s-r)}(i+r)\left(\tilde{T}^{k+1}_{s}(i)-\tilde{T}^{k+1}_{s}(i+1)\right). \nonumber  
\end{eqnarray}
Observe that due to (\ref{condition}), 
\begin{equation}
\mathcal{F}_r^{(k, s)}(i)\equiv 0,\;\; \forall\; 2r\geq s+k+1
\label{forall}
\end{equation}

So, taking into account  proposition \ref{pro:3} and (\ref{forall}), in the case $s+k=2g+1, g\geq k$, we can describe  the  first integrals for  $(k, s)$th equation (\ref{16}) in the following way. Let $\mathcal{F}^{(k, s)}(i)\equiv \left(\mathcal{F}_0^{(k, s)}(i),\ldots, \mathcal{F}_g^{(k, s)}(i)\right)^T$ and 
\begin{equation}
F^{(k, s)}(i)=\left(F_0^{(k, s)}(i), F_0^{(k-1, s-1)}(i+1)\ldots, F_0^{(0, s-k)}(i+k), 0,\ldots, 0\right)^T
\label{vector}
\end{equation}
are $(g+1)$-dimensional vectors and
\[
\mathcal{P}_{s+k}(i)\equiv\left(
\begin{array}{ccccc}
1 & 0 & 0 & \cdots & 0\\
P^1_{s+k}(i) & 1 & 0 & \cdots & 0\\
P^2_{s+k-1}(i) & P^1_{s+k-2}(i+1) & 1 & \cdots & 0\\
P^3_{s+k-2}(i) & P^2_{s+k-3}(i+1) &  P^1_{s+k-4}(i+2) & \cdots & 0\\
\vdots & \vdots & \vdots &  \\
P^g_{s+k-g+1}(i) & P^{g-1}_{s+k-g}(i+1) & P^{g-2}_{s+k-g-1}(i+2) & \cdots & 1
\end{array}
\right).
\]
Then the first integrals are given by the system
\begin{equation}
\mathcal{F}^{(k, s)}(i)=\mathcal{P}_{s+k}(i)F^{(k, s)}(i).
\label{system}
\end{equation}


Respectively, in the case $s+k=2g+2, g\geq k$, we take $(g+2)$-dimensional vectors: $\mathcal{F}^{(k, s)}(i)\equiv \left(\mathcal{F}_0^{(k, s)}(i),\ldots, \mathcal{F}_{g+1}^{(k, s)}(i)\right)^T
$ and $F^{(k, s)}(i)$ given by (\ref{vector}). Then the first integrals are given by system (\ref{system}) with corresponding  matrix
\begin{equation}
\mathcal{P}_{s+k}(i)\equiv\left(
\begin{array}{ccccc}
1 & 0 & 0 & \cdots & 0\\
P^1_{s+k}(i) & 1 & 0 & \cdots & 0\\
P^2_{s+k-1}(i) & P^1_{s+k-2}(i+1) & 1 & \cdots & 0\\
P^3_{s+k-2}(i) & P^2_{s+k-3}(i+1) &  P^1_{s+k-4}(i+2) & \cdots & 0\\
\vdots & \vdots & \vdots &  \\
P^{g+1}_{s+k-g}(i) & P^{g}_{s+k-g-1}(i+1) & P^{g-1}_{s+k-g-2}(i+2) & \cdots & 1
\end{array}
\right).
\label{g+1}
\end{equation}

Let us define now the first integrals $H_r^{(k, s)}(i)\equiv \mathcal{F}_r^{(k, s)}(i)/F_0^{(k, s)}(i)$. Clearly, these integrals are described by the system $H^{(k, s)}(i)=\mathcal{P}_{s+k}(i)\eta^{(k, s)}(i)$  with $\eta^{(k, s)}(i)=\left(1, \eta_1^{(k, s)}(i),\ldots, \eta_k^{(k, s)}(i), 0,\ldots, 0\right)^T$, where $\eta_r^{(k, s)}(i)\equiv F_0^{(k-r, s-r)}(i+r)/F_0^{(k, s)}(i)$ is given in more explicit form by (\ref{eta1}). Notice that making use of relationship (\ref{eta}) we construct a number of the first integrals for $(k, s)$th equation (\ref{17}). As was mentioned above, substituting (\ref{eta}) into (\ref{Hj}), for $r=1,\ldots, k$, produces identities, that is, 
\[
H_r^{(k, s)}(i)=H_r,\;\; r=1,\ldots, k, 
\] 
while doing that for $r\geq k+1$ gives nontrivial polynomial integrals $\{H_{k+1}^{(k, s)}(i),\ldots, H_g^{(k, s)}(i)\}$ in the case  $s+k=2g+1$ and $\{H_{k+1}^{(k, s)}(i),\ldots, H_{g+1}^{(k, s)}(i)\}$ in the case  $s+k=2g+2$, respectively. It should be noted that we can represent these integrals through the relation 
\begin{eqnarray}
R_g(z)&=&z^g+\sum_{j=1}^gH_j^{(k, s)}(i)z^{g-j}\nonumber\\
      &=&\sum_{q=0}^k(-1)^q\tilde{S}^q_{s+k-q+1}(i)\left(z^{g-q}+\sum_{j=1}^{g-q}P^j_{s+k-2q-j+1}(i+q)z^{g-q-j}\right)\nonumber
			\end{eqnarray}
in the case $s+k=2g+1$ and
\begin{eqnarray}
R_{g+1}(z)&=&z^{g+1}+\sum_{j=1}^{g+1}H_j^{(k, s)}(i)z^{g-j+1}\nonumber\\
      &=&\sum_{q=0}^k(-1)^q\tilde{S}^q_{s+k-q+1}(i)\left(z^{g-q+1}+\sum_{j=1}^{g-q+1}P^j_{s+k-2q-j+1}(i+q)z^{g-q-j+1}\right)\nonumber
			\end{eqnarray}
in the case $s+k=2g+2$, respectively. This representation makes our construction of the first integrals more closer to the approach using Lax representation which we have expounded in section \ref{sect4}.

\subsection{Examples}

\subsubsection{Equation (\ref{17}) in the case $k=1$ and $s=3$. }
In this case equation (\ref{17}) is 
\begin{eqnarray}
&&T(i+1)\left(T(i)+T(i+1)+T(i+2)-H_1\right)\nonumber\\
&&\;\;\;\;\;\;\;\;\;\;=T(i+3)\left(T(i+2)+T(i+3)+T(i+4)-H_1\right)\nonumber
\end{eqnarray}
which generate a map
\[
(y_0, y_1, y_2, y_3)\mapsto(y_1, y_2, y_3, \frac{y_1}{y_3}(y_0+y_1+y_2-H_1)-y_2-y_3+H_1).
\]
Polynomial integrals for this map are\footnote{Here discrete polynomials are replaced by multi-variate ones via replacement $T(i+k)\rightarrow y_k$. The notation like $Q^{r, k}_s$ means $k$-shifted polynomials $Q^{k}_s$ corresponding to a shift $i\rightarrow i+1$.}
\[
G_0^{(1, 3)}=\tilde{S}^1_4y_1y_2=\left(y_0+y_1+y_2+y_3-H_1\right)y_1y_2,
\]
\[
\mathcal{G}_1^{(1, 3)}=G^{(0, 4)}_0+Q^{1, 1}_2G^{(1, 3)}_0
=y_0y_1y_2y_3+\left(y_1+y_2\right)\left(y_0+y_1+y_2+y_3-H_1\right)y_1y_2,
\]
\[
\mathcal{F}_2^{(1, 3)}=P^2_3-P^{1, 1}_2\tilde{S}^1_4
=y_0y_2+y_1y_3-\left(y_1+y_2\right)\left(y_0+y_1+y_2+y_3-H_1\right).
\]

\subsubsection{Equation (\ref{16}) in the case $k=1$ and $s=3$}
Equation (\ref{16}) in this case looks as
\[
T(i+4)=T(i)\frac{T(i+2)+T(i+3)+c_1}{T(i+1)+T(i+2)+c_1},
\]
which generate a map
\[
(y_0, y_1, y_2, y_3)\mapsto\left(y_1, y_2, y_3, y_0\frac{y_2+y_3+c_1}{y_1+y_2+c_1}\right).
\]
Using (\ref{system}) with matrix (\ref{g+1}) gives us the following rational integrals for this map:
\[
F_0^{(1, 3)}=\frac{\tilde{T}^{1, 1}_{2}}{y_0y_1y_2y_3}=\frac{y_1+y_2+c_1}{y_0y_1y_2y_3},
\]
\[
\mathcal{F}_1^{(1, 3)}=P^1_4F_0^{(1, 3)}+F_0^{(0, 2),1}=\left(y_0+y_1+y_2+y_3\right)\frac{y_1+y_2+c_1}{y_0y_1y_2y_3}+\frac{1}{y_1y_2},
\]
\[
\mathcal{F}_2^{(1, 3)}=P^2_3F_0^{(1, 3)}+P^{1, 1}_2F_0^{(0, 2), 1}=\left(y_0y_2+y_1y_3\right)\frac{y_1+y_2+c_1}{y_0y_1y_2y_3}+\frac{y_1+y_2}{y_1y_2}.
\]


\section{Discussion}
We  have presented in this paper the way to construct the first integrals for two $(1, 1)$-classes of ordinary difference equations (\ref{122}) and (\ref{132}) which possess Lax pair representation. These equations are presented in terms of special classes of discrete polynomials and it is natural that their first integrals are also should be expressed in terms of these polynomials. Based on conjecture \ref{lem:1}, we have shown the equivalency of dynamical systems generated by two different at the first glance $(k, s)$th equations (\ref{122}) and (\ref{132}) with fixed $(k, s)$. We do not discuss in the paper the Lioville-Arnold integrability for these classes of equations and leave this problem  for further investigation. 
It is also a problem to be addressed to expand  results presented in this paper to general $(h, n)$-classes of difference equations (\ref{12}) and (\ref{13}).   


\section*{Acknowledgments}
I wish to thank the referees for carefully reading the manuscript and for remarks which enabled the presentation of the paper to be improved. This work was partially supported by grant NSh-5007.2014.9. 

\begin{appendix}

\section{Proof of Lemma \ref{lem0}}

Let us prove, for example, (\ref{ST2}) for any odd $r$. To this aim, it is more convenient to present polynomials $S^r_s$ in equivalent form as 
\[
S^r_s(i)=\sum_{\{\lambda_j\}\in \mathcal{B}_{r, s}} T(i+\lambda_1h) T(i+\lambda_{2}h+n)\cdots T(i+\lambda_rh+(r-1)n),
\]
where $\mathcal{B}_{r, s}\equiv \{\lambda_j : 0\leq \lambda_r\leq\cdots\leq\lambda_1\leq s-1\}$ and rewrite (\ref{ST2}) as
\begin{eqnarray}
&&\sum_{j=0}^{(r-1)/2}S^{r-2j}_s(i)T^{2j}_s(i+(r-2j)n)\nonumber\\ 
&&\;\;\;\;\;\;=\sum_{j=0}^{(r-1)/2}S^{r-2j-1}_s(i)T^{2j+1}_s(i+(r-2j-1)n).\label{id1}
\end{eqnarray}
Now we observe the following. It is evident that the product $S^l_s(i)T^{q}_s(i+ln)$ is given by the summation
\begin{equation}
\sum_{\{\lambda_j\}} T(i+\lambda_1h) T(i+\lambda_{2}h+n)\cdots T(i+\lambda_{l+q}h+(l+q-1)n)
\label{summation}
\end{equation}
over the set 
\[
\mathcal{B}_{l, s} * D_{q, s}:=\{\lambda_j : 0\leq \lambda_l\leq\cdots \leq\lambda_1 \leq s-1,\;  0\leq\lambda_{l+1}<\cdots <\lambda_{l+q}\leq s-1 \}. 
\]
One can easily check that 
\begin{equation}
\mathcal{B}_{l, s} * D_{q, s}=K_{l+q, l, s}\sqcup K_{l+q, l-1, s},
\label{id2}
\end{equation}
where 
\[
K_{k, l, s}:= \{\lambda_j : 0\leq \lambda_{l+1}\leq\cdots \leq\lambda_1 \leq s-1,\;  0\leq\lambda_{l+1}<\cdots <\lambda_{k}\leq s-1 \}.
\]
Therefore the left-hand side of relation (\ref{id1}) presents a summation (\ref{summation}) over the set
\[
\mathcal{B}_{r, s}\sqcup\left(\mathcal{B}_{r-2, s} * D_{2, s}\right)\sqcup\cdots\sqcup\left(\mathcal{B}_{1, s} * D_{r-1, s}\right).
\] 
In turn, the right-hand 
side of this relation is given by a summation over the set 
\[
(\mathcal{B}_{r-1,s}* D_{1, s})\sqcup\left(\mathcal{B}_{r-3, s} * D_{3, s}\right)\sqcup\cdots\sqcup D_{r, s}.
\] 
To prove (\ref{id1}), it remains only to notice that, by virtue of (\ref{id2}),  these two sets coincide and look as $\bigsqcup_{j=0}^{r-1} K_{r, j, s}$. For all other cases, the scheme of proof is about the same one.
$\opensquare$

\section{Proof of Lemma \ref{lem1}}

We can easily prove (\ref{Bk})  by adding to the left and right sides of this set-theoretic equality the following  set:
\begin{equation}
\left\{\lambda_j :  \lambda_1=-1\; -1\leq\lambda_2\leq\cdots\leq\lambda_{k}\leq s-1\right\}.
\label{left}
\end{equation}
On the one hand, it is obvious that
\begin{eqnarray}
&&\left\{\lambda_j : \lambda_1=-1,\; -1\leq\lambda_2\leq\cdots\leq\lambda_{k-1}\leq s,\;\lambda_k=s\right\}\nonumber \\
&&\;\;\;\;\;\;\;\;\;\;\sqcup \left\{\lambda_j :  \lambda_1=-1,\; -1\leq\lambda_2\leq\cdots\leq\lambda_{k}\leq s-1\right\}\nonumber \\
&&\;\;\;\;\;\;\;\;\;\;=\left\{\lambda_j : \lambda_1=-1,\; -1\leq\lambda_2\leq\cdots\leq\lambda_{k}\leq s\right\}\nonumber 
\end{eqnarray}
and
\[
\left\{\lambda_j : \lambda_1=-1,\; -1\leq\lambda_2\leq\cdots\leq\lambda_{k}\leq s\right\}\sqcup B_{k, s+1}=B_{k, s+2}^{-},
\]
that is, in a result of adding (\ref{left}) to the left-hand side of (\ref{Bk}), we obtain $B_{k, s+2}^{-}$. On the other hand, we get 
\[
\left\{\lambda_j :  \lambda_1=-1,\; -1\leq\lambda_2\leq\cdots\leq\lambda_{k}\leq s-1\right\}\sqcup B_{k, s}=B_{k, s+1}^{-}\nonumber \\
\]
and
\[
B_{k, s+1}^{-}\sqcup \left\{\lambda_j : -1\leq\lambda_1\leq\cdots\leq\lambda_{k-1}\leq s,\;\lambda_k=s \right\}=B_{k, s+2}^{-}.
\]
Since we obtain in the right-hand side of (\ref{Bk}) the same set as in the left-hand side, this means that (\ref{Bk}) is indeed valid. The similar reasonings are used to prove (\ref{Bk1}).  Namely, we first add  the set
\[
\left\{\lambda_j : 0\leq\lambda_1\leq\cdots\leq\lambda_{k-1}\leq s,\; \lambda_k=s\right\}
\]
to the left and right side of equality (\ref{Bk1}) and then we prove that in a result in both sides of this equality  we obtain the same set $B_{k, s+2}^{-}$.
Therefore we proved this lemma. \opensquare

\section{Proof of Lemma \ref{lem:2}}

To begin with, we notice that relation (\ref{sjk}) for $s=1$ takes the form  
\begin{equation}
S^{k}_1(i)=\prod_{q=0}^{k-1}T(i+q),\;\;\forall k\geq 1.
\label{Sk1}
\end{equation}
It is obvious that the latter is valid simply by virtue of the definition of the polynomials $S^k_s$. Suppose now we have already proved (\ref{sjk}) for some value of $s\geq 1$ and all $k\geq s$. Then  using identities of the form (\ref{s1}) and (\ref{s2}), we get
\begin{eqnarray}
S^{s+1}_{s+1}(i)&=&S^{s+1}_s(i+1)+T(i+s)S^{s}_{s+1}(i) \nonumber \\
&=&S^{s-1}_{s+2}(i+1)T(i+s)T(i+s+1)+T(i+s)S^{s}_{s+1}(i) \nonumber \\
&=&\left(S^{s-1}_{s+2}(i+1)T(i+s+1)+S^{s}_{s+1}(i)\right)T(i+s) \nonumber \\
&=&S^{s}_{s+2}(i)T(i+s), \label{Sk3}
\end{eqnarray}
for this value of $s$ and
\begin{eqnarray}
S^{k+1}_{s+1}(i)&=&S^{k+1}_s(i+1)+T(i+k)S^{k}_{s+1}(i) \nonumber \\
&=&S_{k+2}^{s-1}(i+1)\prod_{q=s}^{k+1}T(i+q)+T(i+k)S_{k+1}^{s}(i)\prod_{q=s}^{k-1}T(i+q) \nonumber \\
&=&\left(T(i+k+1)S_{k+2}^{s-1}(i+1)+S_{k+1}^{s}(i)\right)\prod_{q=s}^{k}T(i+q) \nonumber \\
&=&S_{k+2}^{s}(i)\prod_{q=s}^{k}T(i+q) \label{Sk2}
\end{eqnarray}
for $k\geq s+1$.
 
So supposing that the relation (\ref{sjk}) is valid for some value of $s$ and all value of $k\geq s$, we prove therefore that it is fulfilled for $s+1$ and all $k\geq s+1$.
Now we are in a position, using (\ref{Sk1}), (\ref{Sk3}) and (\ref{Sk2}),  to prove (\ref{sjk}) by induction on $k$ for all values of $s\geq 1$ and $k\geq s$. Therefore this lemma is proved.
\opensquare

\section{Proof of Lemma \ref{lemm:2}}

We first observe that 
\begin{equation}
g_0^s(i)=\prod_{q=0}^{s-1}T(i+q)
\label{g0}
\end{equation}
and
\begin{eqnarray}
g_0^s(i)+f_1^s(i)&=&\prod_{q=0}^{s-1}T(i+q)+\tilde{Q}^1_{s-2}(i+1)\prod_{q=0}^{s-2} T(i+q)\nonumber \\
&=&\left(T(i+s-1)+\tilde{Q}^1_{s-2}(i+1)\right)\prod_{q=0}^{s-2} T(i+q)\nonumber \\
&=&\tilde{S}^1_{s-1}(i+1)\prod_{q=0}^{s-2} T(i+q).\label{f0} 
\end{eqnarray}
Therefore (\ref{1111}) and (\ref{2222}) are fulfilled for $r=0$.
Let us prove this lemma by induction on $r$. We first suppose that (\ref{1111}) is valid for some $r$, then, taking into account an identity of the form (\ref{s3}), we calculate
\begin{eqnarray}
\fl
\left(\sum_{j=0}^rg_{j}^s(i)+\sum_{j=1}^{r+1}f_{j}^s(i)\right)+g_{r+1}^s(i)&=&\tilde{S}^r_{s-r}(i)\tilde{S}^{r+1}_{s-r-1}(i+1)\prod_{q=r}^{s-r-2}T(i+q) \nonumber \\
\fl
&&+\tilde{Q}^{r+1}_{s-r-2}(i+1)\tilde{S}^{r+1}_{s-r-1}(i+1)\prod_{q=r+1}^{s-r-2}T(i+q) \nonumber \\
\fl
&=&\left(T(i+r)\tilde{S}^r_{s-r}(i)+\tilde{Q}^{r+1}_{s-r-2}(i+1)\right)  \nonumber \\
\fl
&&\times\tilde{S}^{r+1}_{s-r-1}(i+1)\prod_{q=r+1}^{s-r-2}T(i+q) \nonumber \\
\fl
&=&\tilde{S}^{r+1}_{s-r-1}(i)\tilde{S}^{r+1}_{s-r-1}(i+1)\prod_{q=r+1}^{s-r-2}T(i+q). \nonumber 
\end{eqnarray}
Therefore we  prove that if (\ref{1111}) is valid for some $r$, then (\ref{2222}) is valid for  $r+1$. Now suppose that  (\ref{2222}) is valid for some $r$, then, using an identity 
of the form (\ref{s444}), we get 
\begin{eqnarray}  
\fl
\left(\sum_{q=0}^rg_{j}^s(i)+\sum_{q=1}^rf_{j}^s(i)\right)+f_{r+1}^s(i)&=&\tilde{S}^r_{s-r}(i)\tilde{S}^{r}_{s-r}(i+1)\prod_{q=r}^{s-r-1}T(i+q) \nonumber \\ 
\fl
&&+\tilde{Q}^{r+1}_{s-r-2}(i+1)\tilde{S}^{r}_{s-r}(i)\prod_{q=r}^{s-r-2}T(i+q) \nonumber \\
\fl
&=&\left(\tilde{Q}^{r+1}_{s-r-2}(i+1)+T(i+s-r-1)\tilde{S}^r_{s-r}(i+1)\right)  \nonumber \\
\fl
&&\times\tilde{S}^{r}_{s-r}(i)\prod_{q=r}^{s-r-2}T(i+q) \nonumber \\
\fl
&=&\tilde{S}^{r+1}_{s-r-1}(i+1)\tilde{S}^{r}_{s-r}(i)\prod_{q=r}^{s-r-2}T(i+q). \nonumber 
\end{eqnarray}
Therefore we  prove that if (\ref{2222}) is valid for some $r$, then (\ref{1111}) is valid for the same  value of $r$. Now, to prove this lemma by induction, it remains to use (\ref{g0}) and (\ref{f0}). Therefore this lemma is proved. \opensquare

\section{Proof of Lemma \ref{lemm:1}}

For further convenience, let us denote 
\[
D_r^{(k, s)}(i)=G_r^{(k, s)}(i)-G_{r+1}^{(k, s-1)}(i).
\]
We first remark that, by definition, 
\begin{eqnarray}
D_{r+1}^{(k, s)}(i)-D_{r}^{(k, s)}(i)&=&Q^{r+1}_{s-k-r-2}(i+k+1)\tilde{S}^{k+r+1}_{s-r}(i)\prod_{q=k+r+1}^{s-r-2}T(i+q)\nonumber \\
                                      &&-Q^{r+2}_{s-k-r-2}(i+k)\tilde{S}^{k+r+1}_{s-r-1}(i)\prod_{q=k+r+1}^{s-r-3}T(i+q).\label{Fr}
\end{eqnarray}
Let us prove this lemma by induction on $r$. Suppose that we  have already proved the validity of (\ref{lemm}) for some $r\geq 0$. Then,  taking into account (\ref{Fr}), we calculate
\begin{eqnarray}
D_{r+1}^{(k, s)}(i)&=&S^r_{s-k-r}(i+k)\tilde{S}^{k+r}_{s-r}(i+1)\prod_{q=k+r}^{s-r-1}T(i+q)\nonumber \\
&&-S^{r+1}_{s-k-r-1}(i+k)\tilde{S}^{k+r}_{s-r}(i)\prod_{q=k+r}^{s-r-2}T(i+q)\nonumber \\
&&+Q^{r+1}_{s-k-r-2}(i+k+1)\tilde{S}^{k+r+1}_{s-r}(i)\prod_{q=k+r+1}^{s-r-2}T(i+q)\nonumber \\
&&-Q^{r+2}_{s-k-r-2}(i+k)\tilde{S}^{k+r+1}_{s-r-1}(i)\prod_{q=k+r+1}^{s-r-3}T(i+q).\nonumber 
\end{eqnarray}
Making use of an identity of the form (\ref{s1}), namely, 
\[
\tilde{S}^{k+r+1}_{s-r}(i)=\tilde{S}^{k+r+1}_{s-r-1}(i+1)+T_{i+k+r}\tilde{S}^{k+r}_{s-r}(i),
\]
we obtain
\begin{eqnarray}
\fl
D_{r+1}^{(k, s)}(i)&=&S^r_{s-k-r}(i+k)\tilde{S}^{k+r}_{s-r}(i+1)\prod_{q=k+r}^{s-r-1}T(i+q)\nonumber \\
\fl
&&+\left(Q^{r+1}_{s-k-r-2}(i+k+1)-S^{r+1}_{s-k-r-1}(i+k)\right)\tilde{S}^{k+r}_{s-r}(i)\prod_{q=k+r}^{s-r-2}T(i+q)\nonumber \\
\fl
&&+Q^{r+1}_{s-k-r-2}(i+k+1)\tilde{S}^{k+r+1}_{s-r-1}(i+1)\prod_{q=k+r+1}^{s-r-2}T(i+q)\nonumber \\
\fl
&&-Q^{r+2}_{s-k-r-2}(i+k)\tilde{S}^{k+r+1}_{s-r-1}(i)\prod_{q=k+r+1}^{s-r-3}T(i+q).\nonumber 
\end{eqnarray}
In turn, using  an identity of the form (\ref{s3}),
we calculate to get
\begin{eqnarray}
\fl
D_{r+1}^{(k, s)}(i)&=&S^r_{s-k-r}(i+k)\left(T_{i+s-r-1}\tilde{S}^{k+r}_{s-r}(i+1)-T_{i+k+r}\tilde{S}^{k+r}_{s-r}(i)\right)\prod_{q=k+r}^{s-r-2}T(i+q)\nonumber \\
\fl
&&+Q^{r+1}_{s-k-r-2}(i+k+1)\tilde{S}^{k+r+1}_{s-r-1}(i+1)\prod_{q=k+r+1}^{s-r-2}T(i+q)\nonumber \\
\fl
&&-Q^{r+2}_{s-k-r-2}(i+k)\tilde{S}^{k+r+1}_{s-r-1}(i)\prod_{q=k+r+1}^{s-r-3}T(i+q)\nonumber\\
\fl
&=&S^r_{s-k-r}(i+k)\left(\tilde{S}^{k+r+1}_{s-r-1}(i+1)-\tilde{S}^{k+r+1}_{s-r-1}(i)\right)\prod_{q=k+r}^{s-r-2}T(i+q)\nonumber \\
\fl
&&+Q^{r+1}_{s-k-r-2}(i+k+1)\tilde{S}^{k+r+1}_{s-r-1}(i+1)\prod_{q=k+r+1}^{s-r-2}T(i+q)\nonumber \\
\fl
&&-Q^{r+2}_{s-k-r-2}(i+k)\tilde{S}^{k+r+1}_{s-r-1}(i)\prod_{q=k+r+1}^{s-r-3}T(i+q)\nonumber\\
\fl
&=& \left(Q^{r+1}_{s-k-r-2}(i+k+1)+T_{i+k+r}S^r_{s-k-r}(i+k)\right)\tilde{S}^{k+r+1}_{s-r-1}(i+1)\prod_{q=k+r+1}^{s-r-2}T(i+q)\nonumber\\
\fl
&&- \left(Q^{r+2}_{s-k-r-2}(i+k)+T_{i+k+r}T_{i+s-r-2}S^r_{s-k-r}(i+k)\right)\prod_{q=k+r+1}^{s-r-3}T(i+q).\nonumber
\end{eqnarray}
Finally, using identities (\ref{s4}) and (\ref{s3}),
we get
\begin{eqnarray}
D_{r+1}^{(k, s)}(i)&=&S^{r+1}_{s-k-r-1}(i+k)\tilde{S}^{k+r+1}_{s-r-1}(i+1)\prod_{q=k+r+1}^{s-r-2}T(i+q)\nonumber\\
&&-S^{r+2}_{s-k-r-2}(i+k)\tilde{S}^{k+r+1}_{s-r-1}(i)\prod_{q=k+r+1}^{s-r-3}T(i+q)\nonumber
\end{eqnarray}
Therefore we prove that if (\ref{lemm}) is valid for some $r\geq 0$ then it is valid for  $r+1$. To prove the lemma, it remains to prove (\ref{lemm}) for $r=0$.
We obtain
\begin{eqnarray}
D_0^{(k, s)}(i)&=&\tilde{S}^k_{s+1}(i)\prod_{q=k}^{s-1}T(i+q)-\tilde{S}^{k-1}_{s+1}(i)\prod_{q=k-1}^{s-1}T(i+q)\nonumber\\
&&-Q^1_{s-k-1}(i+k)\tilde{S}^{k}_{s}(i)\prod_{q=k}^{s-2}T(i+q)\nonumber\\
&=&\left(\tilde{S}^k_{s}(i+1)+T_{i+k-1}\tilde{S}^{k-1}_{s+1}(i)\right)\prod_{q=k}^{s-1}T(i+q)\nonumber\\
&&-\tilde{S}^{k-1}_{s+1}(i)\prod_{q=k-1}^{s-1}T(i+q)-Q^1_{s-k-1}(i+k)\tilde{S}^{k}_{s}(i)\prod_{q=k}^{s-2}T(i+q)\nonumber\\
&=&\tilde{S}^k_{s}(i+1)\prod_{q=k}^{s-1}T(i+q)-S^1_{s-k-1}(i+k)\tilde{S}^{k}_{s}(i)\prod_{q=k}^{s-2}T(i+q).\nonumber
\end{eqnarray}
Therefore this lemma is proved. \opensquare

\section{Proof of Proposition \ref{lemm1}}

Firstly, remark that the relation (\ref{l1}) $=$ (\ref{l2}) is valid by virtue of identities for these polynomials. For $r=0$, relations (\ref{l1}) and
(\ref{l2}) are obvious. To prove the validity of them by induction on $r$, it remains to show the following. Suppose that we have already proved  (\ref{l2}) for some value of $r$, then making use of recurrent relation (\ref{rec}) and identities of the form  (\ref{s3}) and (\ref{s444}), we obtain
\begin{eqnarray}
\fl
G_{r+1}^{(k, s)}(i+1)-G_{r+1}^{(k, s)}(i)&=&\Lambda_{r}^{(k, s)}(i)\left(\tilde{S}^{k+r+1}_{s-r}(i+1)-\tilde{S}^{k+r+1}_{s-r}(i)\right) \nonumber \\ 
\fl
&&+Q^{r+1}_{s-k-r-2}(i+k+2)\tilde{S}^{k+r+1}_{s-r}(i+1)\prod_{q=k+r+2}^{s-r-1}T(i+q) \nonumber \\
\fl
&&-Q^{r+1}_{s-k-r-2}(i+k+1)\tilde{S}^{k+r+1}_{s-r}(i)\prod_{q=k+r+1}^{s-r-2}T(i+q) \nonumber \\
\fl
&=&\left(Q^{r+1}_{s-k-r-2}(i+k+2)+T(i+k+r+1)\right. \nonumber \\
\fl
&&\left.\times S^{r}_{s-k-r}(i+k+1)\right)\tilde{S}^{k+r+1}_{s-r}(i+1)\prod_{q=k+r+2}^{s-r-1}T(i+q) \nonumber \\
\fl
&&-\left(Q^{r+1}_{s-k-r-2}(i+k+1)+T(i+s-r-1) \right. \nonumber \\
\fl
&&\left.\times S^{r}_{s-k-r}(i+k+1)\right)\tilde{S}^{k+r+1}_{s-r}(i)\prod_{q=k+r+1}^{s-r-2}T(i+q) \nonumber \\
\fl
&=&S^{r+1}_{s-k-r-1}(i+k+1)\tilde{S}^{k+r+1}_{s-r}(i+1)\prod_{q=k+r+2}^{s-r-1}T(i+q) \nonumber \\
\fl
&& -S^{r+1}_{s-k-r-1}(i+k+1)\tilde{S}^{k+r+1}_{s-r}(i)\prod_{q=k+r+1}^{s-r-2}T(i+q) \nonumber \\
\fl
&=&\Lambda_{r+1}^{(k, s)}(i)\left(T(i+s-r-1)\tilde{S}^{k+r+1}_{s-r}(i+1) \right. \nonumber \\
\fl
&&\left. -T(i+k+r+1)\tilde{S}^{k+r+1}_{s-r}(i)\right). \nonumber
\end{eqnarray}
Therefore this proposition is proved.    \opensquare 

\section{Proof of Proposition \ref{pro:3}}

Firstly, remark that the relation (\ref{l11}) $=$ (\ref{l21}) is valid by virtue of identities for these polynomials. For $r=0$, relations (\ref{l11}) and
(\ref{l21}) are obvious. To prove the validity of them by induction on $r$, it remains to show the following. Suppose that we have already proved  (\ref{l21}) for some value of $r$, then making use of recurrent relation (\ref{rec1}), we obtain 
\begin{eqnarray}
\fl
F_{r+1}^{(k, s)}(i+1)-F_{r+1}^{(k, s)}(i)&=&\Delta_{r}^{(k, s)}(i)\left(\tilde{T}^{k+r+1}_{s+r}(i-r)-\tilde{T}^{k+r+1}_{s+r}(i-r+1)\right) \nonumber \\ 
\fl
&&+P^{r+1}_{s+k+r+2}(i-r)\frac{\tilde{T}^{k+r+1}_{s+r}(i-r+1)}{\prod_{q=-r}^{s+k+r+1}T(i+q)} \nonumber \\
\fl
&&-P^{r+1}_{s+k+r+2}(i-r-1)\frac{\tilde{T}^{k+r+1}_{s+r}(i-r)}{\prod_{q=-r-1}^{s+k+r}T(i+q)} \nonumber \\
\fl
&=&\left(\frac{P^{r+1}_{s+k+r+2}(i-r)}{T(i+k+s+r+1)}-T^r_{s+k+r}(i-r+1)\right)\frac{\tilde{T}^{k+r+1}_{s+r}(i-r+1)}{\prod_{q=-r}^{s+k+r}T(i+q)} \nonumber \\
\fl
&& -\left(\frac{P^{r+1}_{s+k+r+2}(i-r-1)}{T(i-r-1)}-T^r_{s+k+r}(i-r+1)\right)\frac{\tilde{T}^{k+r+1}_{s+r}(i-r)}{\prod_{q=-r}^{s+k+r}T(i+q)}.\nonumber 
\end{eqnarray}
By virtue of identities (\ref{PT1}) and (\ref{PT2}), we get
\begin{eqnarray}
\fl
F_{r+1}^{(k, s)}(i+1)-F_{r+1}^{(k, s)}(i)&=&\frac{T^{r+1}_{s+k+r+1}(i-r)}{\prod_{q=-r}^{s+k+r}T(i+q)}\left(\frac{\tilde{T}^{k+r+1}_{s+r}(i-r+1)}{T(i+k+s+r+1)}-\frac{\tilde{T}^{k+r+1}_{s+r}(i-r)}{T(i-r-1)}\right)\nonumber \\
\fl
&=&\Delta_{r+1}^{(k, s)}(i)\left(T(i-r-1)\tilde{T}^{k+r+1}_{s+r}(i-r+1)\right.\nonumber \\
\fl
&&\left.-T(i+k+s+r+1)\tilde{T}^{k+r+1}_{s+r}(i-r)\right).\nonumber 
\end{eqnarray}
Therefore this proposition is proved.    \opensquare

\section{Proof of Conjecture \ref{lem:1} for $r=2$ and $r=3$}

Unfortunately, at the moment, we can not prove identity (\ref{identity1})  for all $r$. Clearly, in the case $r=1$, it is evident. Below we give a proof of this conjecture  for $r=2, 3$.

\subsection{The case $r=2$}
We start from the proven identity of the form (\ref{ST21}), namely,
\begin{equation}
\left(S, T\right)^2_s(i):=S^2_s(i)-S^1_s(i)T^1_s(i+1)+T^2_s(i)=0,\;\; \forall s\geq 1
\label{S^2}
\end{equation}
and then we first transform (\ref{S^2}) to
\[
S^2_s(i)-\left(S^1_{s+1}(i)-T(i+s)\right)T^1_s(i+1)+T^2_s(i)=0
\]
or
\[
S^2_s(i)-S^1_{s+1}(i)T^1_s(i+1)+T^2_s(i)=-T(i+s)T^1_s(i+1)
\]
an then to
\[
S^2_s(i)-S^1_{s+1}(i)\left(T^1_{s-1}(i+1)+T(i+s)\right)+T^2_s(i)=-T(i+s)T^1_s(i+1)
\]
or
\begin{eqnarray}
\left\langle S, T\right\rangle^2_s(i) &=&T(i+s)\left(S^1_{s+1}(i)-T^1_s(i+1)\right)\nonumber\\
&=&T(i)T(i+s).\nonumber
\end{eqnarray}

\subsection{The case $r=3$}

We start from the proven identity
\begin{eqnarray}
\left(S, T\right)^3_s(i)&\equiv& S^3_s(i)-S^2_s(i)T^1_s(i+2)+S^1_s(i)T^2_s(i+1)-T^3_s(i)\nonumber\\
&=&0,\;\; \forall s\geq 1.
\label{S^3}
\end{eqnarray}
Making use of the identities  of the form (\ref{s2}) we rewrite  (\ref{S^3}) in equivalent form
\begin{eqnarray}
&& S^3_s(i)-S^2_s(i)T^1_s(i+2)+S^1_s(i)T^2_s(i+1)-T^3_s(i)\nonumber\\
                               &&\;\;= S^3_s(i)-\left(S^2_{s+1}(i)-T(i+s)S^1_{s+1}(i+1)\right)T^1_s(i+2)\nonumber\\
                               & &\;\;\;+\left(S^1_{s+2}(i)-T(i+s)-T(i+s+1)\right)T^2_s(i+1)-T^3_s(i)\nonumber
\end{eqnarray}
or
\begin{eqnarray}
&& S^3_s(i)-S^2_{s+1}(i)T^1_s(i+2)+S^1_{s+2}(i)T^2_s(i+1)-T^3_s(i)\nonumber\\
                                      &&\;\;=T(i+s)\left(T^2_s(i+1)-T^1_s(i+2)S^1_{s+1}(i+1)\right)+T(i+s+1)T^2_s(i+1).\nonumber
\end{eqnarray}
Next, using an identities of the form (\ref{t222}), we transform the latter into\footnote{It is evident that the identity $T^1_s(i+2)=T^1_{s-2}(i+2)+T(i+s)+T(i+s+1)$ is valid only for $s\geq 2$.}
\begin{eqnarray}
&&S^3_s(i)-S^2_{s+1}(i)\left(T^1_{s-2}(i+2)+T(i+s)+T(i+s+1)\right)\nonumber\\
                                     &&+S^1_{s+2}(i)\left(T^2_{s-1}(i+1)+T(i+s+1)T^1_{s-1}(i+1)\right)-T^3_s(i)\nonumber
\end{eqnarray}
or
\begin{eqnarray}
\left\langle S, T\right\rangle^3_s(i) &\equiv&T(i+s)\xi_{2, 0}(i)+T(i+s+1)\xi_{2, 1}(i) \nonumber
\end{eqnarray}
with
\[
\xi_{2, 0}(i)\equiv T^2_s(i+1)-T^1_s(i+2)S^1_{s+1}(i+1)+S^2_{s+1}(i)
\]
and
\[
\xi_{2, 1}(i)\equiv T^2_s(i+1)-T^1_{s-1}(i+1)S^1_{s+2}(i)+S^2_{s+1}(i).
\]
It remains to convert $\xi_{2, 0}$ and $\xi_{2, 1}$ into desired form. Using the identities (\ref{s1}) and (\ref{s2}), we get
\begin{eqnarray}
\fl
\xi_{2, 0}(i)&=&T^2_s(i+1)-T^1_s(i+2)\left(S^1_{s}(i+1)+T(i+s+1)\right)\nonumber\\
\fl
             & &+S^2_{s}(i+1)+T(i+1)S^1_{s+1}(i)\nonumber\\
						\fl
						 &=& T(i+1)S^1_{s+1}(i)-T(i+s+1)T^1_s(i+2)\nonumber\\
						\fl
						 &=& T(i+1)\left(S^1_{s+2}(i)-T(i+s+1)\right)-T(i+s+1)\left(T^1_{s+1}(i+1)-T(i+1)\right)\nonumber\\
						\fl
						 &=& T(i+1)S^1_{s+2}(i)-T(i+s+1)T^1_{s+1}(i+1)\nonumber
\end{eqnarray}
and
\begin{eqnarray}
\fl
\xi_{2, 1}(i)&=&T^2_s(i+1)-T^1_{s-1}(i+1)\left(S^1_{s}(i+2)+T(i)+T(i+1)\right)\nonumber\\
\fl
             & &+S^2_{s}(i+1)+T(i+1)S^1_{s+1}(i)\nonumber\\
						\fl
						&=&T^2_s(i+1)-\left(T^1_{s}(i+1)-T(i+s)\right)S^1_{s}(i+2)+S^2_{s}(i+1)\nonumber\\
						\fl
						&&+T(i+1)\left(T(i)+T(i+s)\right)-T(i)T^1_{s-1}(i+1)\nonumber\\
						\fl
						&=&-T(i)\left(T^1_{s-1}(i+1)-T(i+1)\right)+T(i+s)\left(S^1_{s}(i+2)+T(i+1)\right)\nonumber\\
						\fl
						&=&-T(i)T^1_{s-2}(i+2)+T(i+s)S^1_{s+1}(i+1).\nonumber
\end{eqnarray}
Therefore
\begin{eqnarray}
\fl
\left\langle S, T\right\rangle^3_s(i)&=&T(i+s)\left(T(i+1)S^1_{s+2}(i)-T(i+s+1)T^1_{s+1}(i+1)\right)\nonumber\\
\fl
&&+T(i+s+1)\left(-T(i)T^1_{s-2}(i+2)+T(i+s)S^1_{s+1}(i+1)\right)\nonumber\\
\fl
&=&T(i+s)T(i+s+1)\left(S^1_{s+1}(i+1)- T^1_{s+1}(i+1)\right)\nonumber\\
\fl
&&+T(i+1)T(i+s)S^1_{s+2}(i)-T(i)T(i+s+1)T^1_{s-2}(i+2)\nonumber\\
\fl
&=&T(i+1)T(i+s)S^1_{s+2}(i)-T(i)T(i+s+1)T^1_{s-2}(i+2).\nonumber
\end{eqnarray}
Therefore we have proved (\ref{identity1}) for $k=3$ and all $s\geq 2$. Moreover, by virtue of (\ref{identity8}), we have 
\[
\left\langle S, T\right\rangle^3_1(i)=S^3_1(i)=T(i)T(i+1)T(i+2).
\]
$\opensquare$


\end{appendix}


\end{document}